%% file: RA_RapIRSA_COMMET-Elsv_ACCEPTED_FINAL.tex
\documentclass[12pt,onecolumn,a4paper, draftcls]{IEEEtran}
\usepackage[utf8]{inputenc}
\usepackage[T1]{fontenc}

\usepackage{amsfonts,amssymb,amsthm}
\usepackage{float}
\usepackage{caption}
\usepackage[ruled,vlined]{algorithm2e}
\usepackage{multirow}
\usepackage{array,makecell,multirow,tabularx,ragged2e,booktabs}
\usepackage{placeins}
\usepackage{subcaption}
\usepackage{balance}
\usepackage{tikz}
\usetikzlibrary{shapes}
\usepackage{lscape}
\usepackage{mathtools}
\usepackage{comment}

\usepackage{lineno}

\newtheorem{defn}{Definition}

\usepackage{graphicx}

\usepackage{amsmath}

\DeclarePairedDelimiter\floor{\lfloor}{\rfloor}

\usepackage{array}
\usepackage{stfloats}

\usepackage{url}
\usepackage{soul} 

\usepackage{tikz}
\definecolor{gold}{rgb}{0.85,.66,0}
\definecolor{cian}{rgb}{.02,.7,.95}
\definecolor{dgreen}{rgb}{0,.4,0}

\newcommand{\ta}{\textcolor{black}}

\definecolor{colororange}{HTML}{E65100} 
\definecolor{colordgray}{HTML}{795548} 
\definecolor{colorhgray}{HTML}{212121} 
\definecolor{colorgreen}{HTML}{009688} 
\definecolor{colorlgray}{HTML}{FAFAFA} 
\definecolor{colorblue}{HTML}{0277BB} 
\definecolor{colorred}{HTML}{DC143C} 

\newcolumntype{L}{>{\centering\arraybackslash}m{3cm}}
\renewcommand{\vec}[1]{\boldsymbol{#1}}

\usepackage{color, colortbl}

\usepackage{framed} 

\usepackage[acronyms,nonumberlist,nogroupskip]{glossaries} 
\usepackage[abbreviations]{glossaries-extra}
\setabbreviationstyle[acronym]{long-short}
\usepackage{glossary-superragged}
 \usepackage{glossary-mcols} 
\newglossary[slg]{symbols}{sym}{sbl}{Symbols}
\makeglossaries
\input{Acronyms}

\begin{document}
\title{Raptor-IRSA Grant-free Random Access Protocol for Smart Grids Applications}

\author{{Angel Esteban Labrador Rivas}, {Taufik Abrão}\\
\thanks{This work was partly supported by the National Council for Scientific and Technological Development (CNPq) of Brazil under Grants 310681/2019-7, and in part by the CAPES (Financial code 001), and by State University of Londrina (UEL), PR, Brazil.}
\thanks{A. E. L. Rivas and T.  Abrão are with the Electrical Engineering Department, State University of Londrina, PR, Brazil.  Rod. Celso Garcia Cid - PR445,  s/n, Campus Universitário, Po.Box 10.011.  CEP: 86057-970. E-mail: \texttt{angel.labrador@uel.br}; \quad \texttt{taufik@uel.br}}
}

\maketitle	
		
\begin{abstract}
This paper deals with the reliability of random access (RA) protocols for massive wireless smart grid communication (m-SGC).  We propose and analyze an improved grant-free RA (GF-RA) protocol for critical SG applications under strict QoS m-SGC requirements. At first, we discuss the main features of the SG neighborhood area network (NAN) architecture. We explore the main features of low-rate machine-type wireless networks, and also we describe a technology characterization of wireless neighborhood area networks (WNAN) in medium-range coverage applications. We propose a new-improved irregular repetition slotted ALOHA, combing Raptor codes and irregular ALOHA, namely RapIRSA random access protocol, to better respond to critical high-reliability QoS requirements under a 5G network perspective. Then, we compare and comprehensively analyze the proposed RapIRSA protocol with two existing RA protocols, the IRSA protocol, and the classical Slotted Aloha. Finally, We summarize the potential challenges in implementing the proposed RA protocol for SG critical applications considering many smart sensors (SS).
\end{abstract}




\section{Introduction} 

Most of the modern \gls{SG} \gls{QoS}  models arrange a mapping method with an electrical-telecommunication design.  The industry application experience and electrical engineering decisions rarely foresaw \gls{SG} service requirements. 

Wireless communications represent a challenge to \gls{SG} implementation as many mission-critical applications require real-time data transfer. Maintaining delay and reliability \gls{SG} requirements over wireless shared network disputes with the possibility to extend the number of sensor devices in a wireless network \cite{Khalid2021}. Also, wireless communications bring several security concerns for \gls{SG}; \ta{authors in \cite{Khalid2021}} explore techniques to improve the robustness of wireless mesh networks for mission-critical applications.	As \gls{SG} states as a cyber-physical grid with \gls{IoT} and {5G} data acquisition integration multiple access emerges as trending research \cite{Shahinzadeh2020}.

A \gls{GB} scheduling operation guarantees that a user has restricted resources to the wireless channel, consequently circumventing unspecified collisions and increasing latency and communication overhead \cite{Liu2020}.  On the other hand, \gls{GF} based \gls{RA} protocols represent a solution to decrease the access latency \cite{Berardinelli2018}. \gls{GF}-based protocols use transmission over shared resources if multiple neighboring users transmit simultaneously. Oversharing creates potential collisions that jeopardize transmission reliability. Currently,  academic research and standardization commissions have proposed approaches to enhance the backed traffic loading with \gls{GF} \gls{RA} schemes while guaranteeing high reliability and low latency \cite{Recayte2020}.

Several ALOHA-like schemes use \gls{SIC} techniques to resolve multiple collided packets \cite{Liva2011,Oinaga2020,Huang2021}.  A certain number of time-slots compose the \gls{Saloha} temporal frame where the user transmits a predefined number of times. \Gls{IRSA} optimizes the probability mass function to maximize the peak throughput performance.  \gls{IRSA} performance \ta{results} better than S-ALOHA and \gls{CRDSA}  \cite{Liva2011,Munari2020,Moroglu2020,Saha2021}. \ta{Furthermore, the number of time-slot adjustments relies on the \gls{BS} tasks since transmission reaches the nominal throughput.} Thus, a proper time-frame length avoids decay in the throughput performance.

\subsection{Related Works} 
Recent literature addressing joint issues involving the problem of data collision, access delay, and/or power consumption in random access (RA) Internet of things (IoT) have been arising \cite{Zhai2022}, \cite{Shang2021}, \cite{Dohler2017}, and\cite{Xie2022}. {\it E.g.,} a random access scheme for large-scale micro-power wireless IoT sensors based on {\it slot-scheduling and hybrid mode} has been proposed recently in \cite{Zhai2022}. This scheme is based on different time-slot structures, applying a specific slot-scheduling procedure according to network workload and power consumption. Sensors with varying service priorities are arranged in other time slots for competitive access using an appropriate RA mechanism. Besides, the algorithm rationally places the number of time slots and competing devices into different time slots. This scheme can meet the timeliness requirements of different services and reduce the overall network power consumption when dealing with different RA scenarios, effectively reducing the overall power consumption. In contrast, high-priority services can meet the promptness requirements based on lower power consumption. 
Raptor-coded random access protocol to enable reliable transmission in the decentralized unmanned aerial vehicle (UAV) network is proposed in \cite{Shang2021}. Since the considered network is composed of several overlapped RA interfering subsystems,  the proposed Raptor-based RA scheme can reduce the bit-error rate (BER) by implementing three steps: {\it i})  selecting the number of time-slots based on a derived lower bound for reliable RA operation; {\it ii}) {\it error-correcting codes} are incorporated as precoding before RA operation; {\it iii}) by correlating two consecutive slots, an idle-slot-filling approach is aggregated to improve the RA system efficiency further. Numerical results reveal that the proposed 3-step strategy substantially reduces the BER while can save a substantial number of slots ($\approx20\%$) compared with the existing frameless ALOHA scheme to attain a specific target BER. {Several machine-type communication (MTC) random access devices can simultaneously transmit in the same resource block by incorporating Raptor codes, significantly reducing the access delay and improving the achievable system throughput \cite{Xie2022}. A simple yet efficient random access strategy is proposed in \cite{Dohler2017}, allowing both to detect the selected preambles and to estimate the number of devices that have chosen them. No device identification is needed in the RA phase, significantly reducing the signaling overhead. The maximum number of supported MTC devices in a resource block (RB) is characterized as a function of message length, available resources, and preambles. The proposed scheme can effectively support many MTC devices for a limited number of available resources when the message size is small.}

{Differently of \cite{Zhai2022}, \cite{Shang2021}, \cite{Dohler2017} and \cite{Xie2022}, in this contribution, we deal with the challenges posed by wireless communications in the Smart Grids (SGs) scenarios and implementation, where many mission-critical applications require real-time data transfer. The challenge is maintaining the delay and reliability requirements in the SG over wireless shared network disputes, with the possibility to extend the number of sensor devices over the wireless network \cite{Khalid2021}. To overcome this challenge, industry application experience and engineering decision-makers must be considered in conceiving new methods under an electrical-telecommunication design approach, not only considering the power-energy design perspective. Moreover, wireless communications also bring several security concerns for SG systems; for instance, authors in \cite{Khalid2021} explore techniques to improve the robustness of wireless mesh networks for mission-critical applications. As SG systems have been conceived as a cyber-physical grid with IoT and 5G data acquisition integration, multiple access communication schemes emerge as trending research inside the SG systems \cite{Shahinzadeh2020}.} 

\subsection{Contributions} The focus of this paper is to analyze the influence of \gls{GF}-RA protocols in achieving \gls{QoS} \gls{m-SGC} requirements and {propose} a RA protocol for critical \gls{SG} applications considering a massive number of \gls{SS}. The main \textit{contributions} of this work are threefold:
\begin{itemize}
\item[\textit{i})] {Elaborating on and proposing a new grant-free protocol that leverages Raptor codes and IRSA to improve the performance of RA protocols for \gls{SG} applications. Our approach modifies the standard IRSA protocol by employing Raptor codes to reduce the number of preamble messages and the associated overhead while improving reliability and latency and adapting the protocol to account for different priority levels and varying numbers of devices in the network.}
\item[\textit{ii})] {Thoroughly characterizing the modifications required to adapt RA protocols to existing applications and varying numbers of devices in \gls{SG} applications. Our work provides comprehensive guidance on the specific modifications required to ensure that RA protocols can achieve optimal performance and meet the \gls{QoS} requirements of various \gls{SG} applications.}
\item[\textit{iii})] {Exploring the \gls{QoS} requirements of critical \gls{SG} applications, such as AMI and load management, and characterizing the requirements of \gls{SGC} systems to provide insights into the trade-offs and performance requirements of different RA protocols.}
\end{itemize}

Overall, our proposed grant-free RA protocol, along with our detailed characterization of the necessary modifications and \gls{QoS} requirements, represent a significant contribution to the field of \gls{SG} communications and provides valuable insights into the design of RA protocols for \gls{m-SGC} networks.

The remainder of the paper is organized as follows. The \gls{m-SGC}  schemes and applications are revisited in Section \ref{sec:QoS}. {Random Access protocols suitable for \gls{m-SGC} systems are considered in Section \ref{sec:RA-SGC}, while} 
Section \ref{sec:QoS_Metrics} {describes important metrics deployed in the evaluation of \gls{QoS} of} SG systems. 
Section \ref{sec:Proposed_RA} introduces the proposed RA protocol. Extensive numerical results are explored in Section \ref{sec:results}.
Section \ref{sec:conclusion} draws the main conclusions and future trends. 

\section{\gls{m-SGC}  Architecture and Applications} \label{sec:QoS}
{Smart grid (SG) applications have been implemented using various network technologies such as power line communication (PLC), wireless mesh networks, and low-power wide area networks (LPWAN) \cite{Gungor2018}. LPWAN has emerged as a popular technology for SG and smart city applications due to its long-range connectivity, low power consumption, and low cost \cite{Centenaro2016}. LPWAN technologies such as LoRaWAN and Sigfox are particularly well-suited for SG applications because they can support many low-cost, low-power devices with infrequent data transmission \cite{Ahmed2017}. Moreover, LPWAN enables the communication between devices that are located in hard-to-reach areas, such as underground or in remote locations \cite{Bourdena2017}. These benefits make LPWAN an attractive choice for SG applications, particularly for use cases involving smart metering, distribution automation, and demand-side management \cite{Centenaro2016}.}

{In addition, one can note that \gls{SS} and \gls{SG} applications typically rely on \gls{IoT} and various network technologies, including LPWAN \cite{Saleem2019,Chaudhari2020,Kaveh2020}. These technologies play a significant role in meeting the evolving requirements of applications and services while providing a framework for enabling robust and dynamic solutions. However, LPWAN is not the only technology used for SG applications, and other network technologies such as PLC and wireless mesh networks have also been used \cite{Gungor2018}.}

\Gls{SUN} targets multiple applications within shared networks \cite{Choi2016}. For a utility, this implies performing monitoring and control over the same resources. \gls{SUN} devices provide wide-long-range point-to-point connections, including many outdoor devices. \gls{SUN} aims for low-power wireless applications that usually demand the maximum transmit power available under proper administration. \gls{SUN} application requirements demand a peer-to-peer topology \cite{Society2020}. 
In a star topology, devices establish communications with a single central controller, named the \gls{NAN} coordinator. {In contrast, each device in a peer-to-peer network can route messages among and through any device, making it more flexible for certain applications.}

Most of the delay allowances discussed in this work are a compilation of the requirements specified in previous works \cite{IEEE1646-2005,Chan2006,ITU2003,ITU-I2014,Hu2010}.  {In \cite{Deshpande2011}, the authors provide a quantitative characterization of priorities for \gls{SG} applications based on these standards, recommendation documents, pragmatic needs for utility, and Table \ref{tab:latency} includes the priority of an application relative to others.}
{Besides, {\it delay}} allowances listed {in this table} 
are end-to-end  Uplink delays.

{Note that the latency requirements for \gls{SUN} applications may vary depending on the specific use case and application. For example, some applications may require strict latency requirements to ensure proper operation, while others may be more tolerant of delay.}

\begin{table}[htbp!]
\centering
\caption{Latency \& Priority by \gls{SG} application type.}
\label{tab:latency}
\resizebox{0.6\textwidth}{!}{%
\begin{tabular}{lcc}
\hline {\textbf{Application type}}  & \textbf{\begin{tabular}[c]{@{}c@{}} Latency \\(ms) \end{tabular}} & \textbf{\begin{tabular}[c]{@{}c@{}}Priority \\ 0-max 100-min\end{tabular}} \\ \hline
\textit{Teleprotection (60 Hz, 50 Hz)} & 8,10 & 10 \\
\textit{SCADA} & 10 & 20 \\
\textit{Teleprotection} & 16 & 15 \\
\textit{Synchrophasors} & 20 & 12 \\
\textit{SCADA} & 100 & 25 \\
\textit{Distribution automation} & 100 & 26 \\
\textit{\begin{tabular}[c]{@{}l@{}}Distributed generation - \\ distributed  storage\end{tabular}} & 100 & 27 \\
\textit{MWF} & 100 & 30 \\
\textit{Business voice} & 200 & 60 \\
\textit{Dynamic Line Rating} & 200 & 28 \\
\textit{CCTV} & 200 & 55 \\
\textit{SCADA, DA, DG/DS, DLR} & 200 & 45 \\
\textit{Business data} & 250 & 70 \\
\textit{AMI} & 250 & 40 \\
\textit{Protection} & 500 & 80 \\
\textit{Many/others} & 2000 & 100 \\ \hline
\multicolumn{3}{l}{\tiny This table shows only higher priority applications by application type (non-exhaustive)}
\end{tabular}%
}
\end{table}

\subsection{{Types of Frame in 5G Mobile networks}}\label{sec:Type_frames}

{In smart grids, the design of wireless communication systems and protocols faces significant challenges due to the heterogeneity of prosumers, nodes that can consume and produce electrical power, and high reliability and latency requirements and data rates. Defining the technical requirements for smart grid applications, such as latency, rates, and reliability, remains an ongoing research area. Recent studies have shown that existing wireless communication technologies are insufficient to meet the stringent demands of a time-critical smart grid with strict requirements on latency \cite{Ma2019}. The typical response time for different SG applications can be found in Table \ref{tab:latency}.  However, there are no specifications on latency for many new scenarios of smart grid, such as nano-grids\footnote{{SG in resident areas that may have complex nodes such as prosumers.}}, which are supposed to have even higher technical standards when connected to main grids. In Table \ref{tab:latency}, different instances of SCADA refer to different sub-types or use cases, such as transmission SCADA, distribution SCADA, and substation automation SCADA, each with unique latency and priority requirements.}

{The 5G systems have been developed for supporting various services, such as enhanced mobile broadband (eMBB), massive machine-type communication (mMTC), and ultra-reliable low-latency communication (URLLC). The latter two constitute \gls{IoT} enablers. The high propagation loss in the above 3 GHz spectrum, the limited number of UL slots in a time-division duplexing (TDD) frame, and the limited user power strongly restrict the wireless communication coverage. Moreover, the stringent requirements of eMBB and IoT applications lead to 5G challenges, including site planning, ensuring seamless coverage, adapting the TDD DL/UL slot ratio, and the frame structure for maintaining a low BER, as well as low latency.  Conflicting requirements of high-transmission efficiency $\times$ large coverage area $\times$ low latency can be balanced by properly exploiting the spectrum, enabling the IoT and eMBB services support.}

{\gls{NR} {\it frame structure} is specified in 3GPP specification (38.211) \cite{NR_PHY-3GPP}. 5G \gls{NR} retains the concept of a 10-millisecond frame divided into ten one-millisecond subframes. There is also a slot concept, but its definition differs from a slot in LTE. One slot is defined as 14 OFDM symbols corresponding to once a frame or one millisecond in LTE frame legacy. A frame has a duration of 10 ms, which consists of 10 subframes having 1ms duration, each similar to LTE technology. Each subframe can support time slots of 2$\mu$s duration. Each slot typically consists of 14 OFDM symbols. The radio frame of 10 ms is transmitted continuously as per TDD topology, one after the other. A subframe is of fixed duration {\it i.e.} 1ms, whereas slot length varies based on subcarrier spacing and the number of slots per subframe.  It is 1 ms for 15 KHz, 500 $\mu$s for 30 KHz, and so on. Subcarrier spacing of 15 KHz occupies 1 slot per subframe, subcarrier spacing of 30 KHz occupy 2 slots per subframe, and so on. Each slot occupies either 14 OFDM symbols or 12 OFDM symbols based on normal \gls{CP} and extended CP respectively \cite{Vihriala2016}.}

\section{RA Protocols for Required \gls{QoS} in \gls{m-SGC} Networks} \label{sec:RA-SGC}

In this section, we define suitable network scenarios for RA protocol applied to \gls{m-SGC} networks. We consider a cellular-based single-cell system with \ta{up to} $M$ total users connected to the \gls{BS}, where \ta{$m(t)$} users are active at any time occurrence $t$\ta{; or more specifically at the $k$-th time-slot, $m(\tau_k)$}. {Furthermore, a Beta or a Poisson distribution determined the distribution of the active users $m$ (see Section \ref{ssec:arrival}).}. 

\subsection{{\textsc{Aloha}-based} Classical RA Protocol for \gls{SG} systems}

\textsc{Aloha}-based protocols are among the RA protocols group of high-reliability and high-throughput applications \cite{Berioli2016}. This work concentrates on extensions of the {\textsc{Aloha}} model, holding terminals send a slotted packet over the channel as quickly as it is produced and without performing distributed coordination strategies.

In the sequence, we define the premises for evaluating the performance of the implemented/developed RA protocols in terms of two metrics. 
{The first metric, {\it throughput} ($S$), summarizes the average amount of data blocks accomplished at the receiver over a reference period. The second is a complementary metric, the {\it packet loss ratio} (PLR), eq. \eqref{eq:PLR}.}

\vspace{2mm}
\subsubsection{Slotted ALOHA} 
\begin{figure}[!htbp]
\small
    \centering
    \begin{tikzpicture}[scale=.75, transform shape]
        \color{black}
        \draw[dashed,step=1cm,black,thin] (-0.5,0.001) grid (9.9,4.99);
        \path (-1,4.5) node(u1) {User 1};
        \path (-1,3.5) node(u1) {User 2};
        \path (-1,2.5) node(u1) {User 3};
        \path (-1,1.6) node(u) {$\vdots$};
        \path (-1,0.5) node(um) {User $m$};
        \draw[thick,<->,black] (0,5.5) -- (9.9,5.5) node[pos=0.5,above] {Frame, $T_F$ seconds, $n$ Slots};
        \draw[thick,<->,black] (7,4.5) -- (8,4.5) node[pos=0.5,above] {Slot, $t_\text{s}$};
        \filldraw[fill=red!60!white, draw=black] (1,4) rectangle (2,5);
        \filldraw[fill=blue!40!white, draw=black] (5,3) rectangle (6,4);
        \filldraw[fill=green!40!white, draw=black] (8,2) rectangle (9,3);
        \filldraw[fill=orange!40!white, draw=black] (8,0) rectangle (9,1);
        
 \draw[thick,red] (8.5,1.5) circle [x radius=1, y radius=2] node {Collision};
    \end{tikzpicture}
    {\bf a}) S-ALOHA protocol\\
    \vspace{3mm}

\begin{tikzpicture}[scale=.75, transform shape]
       \color{black}
        \draw[dashed,step=1cm,black,thin] (-0.5,0.001) grid (9.9,4.99);
        \path (-1,4.5) node(u1) {User 1};
        \path (-1,3.5) node(u1) {User 2};
        \path (-1,2.5) node(u1) {User 3};
        \path (-1,1.6) node(u) {$\vdots$};
        \path (-1,0.5) node(um) {User $m$};
        \draw[thick,<->,black] (0,5.5) -- (9.9,5.5) node[pos=0.5,above] {\gls{IRSA} \gls{RAF} Frame, $T_{RAF}$ seconds, $n_{RAF}$ Slots};
        \filldraw[fill=red!60!white, draw=black] (1,4)  rectangle (2,5) node[pos=0.5] {RP 1};
        \filldraw[fill=red!60!white, draw=black] (3,4)  rectangle (4,5) node[pos=0.5] {RP 2};
        \filldraw[fill=red!60!white, draw=black] (5,4)  rectangle (6,5) node[pos=0.5] {RP 3};
        \filldraw[fill=blue!40!white, draw=black] (5,3) rectangle (6,4) node[pos=0.5] {RP 1};
        \filldraw[fill=blue!40!white, draw=black] (8,3) rectangle (9,4) node[pos=0.5] {RP 2};
        \filldraw[fill=green!40!white, draw=black] (3,2) rectangle (4,3) node[pos=0.5] {RP 1};
        \filldraw[fill=green!40!white, draw=black] (8,2) rectangle (9,3) node[pos=0.5] {RP 2};
        \filldraw[fill=orange!40!white, draw=black] (6,0) rectangle (7,1) node[pos=0.5] {RP 1};
        \filldraw[fill=orange!40!white, draw=black] (8,0) rectangle (9,1) node[pos=0.5] {RP 2};

\draw[thick,red] (3.5,3.5) circle [x radius=1, y radius=2] node {Collision};
        \draw[thick,red] (5.5,4) circle [x radius=1, y radius=1.5] node[right] {Collision};
        \draw[thick,red] (8.5,2) circle [x radius=1, y radius=2] node[left] {Collision};
    \end{tikzpicture}
   {\bf b}) \gls{IRSA} protocol
    \caption{Sketch of a) S-ALOHA; b) \gls{IRSA} protocols as presented by \cite{Liva2011}}
    \label{fig:S-ALOHA_IRSA}
\end{figure}
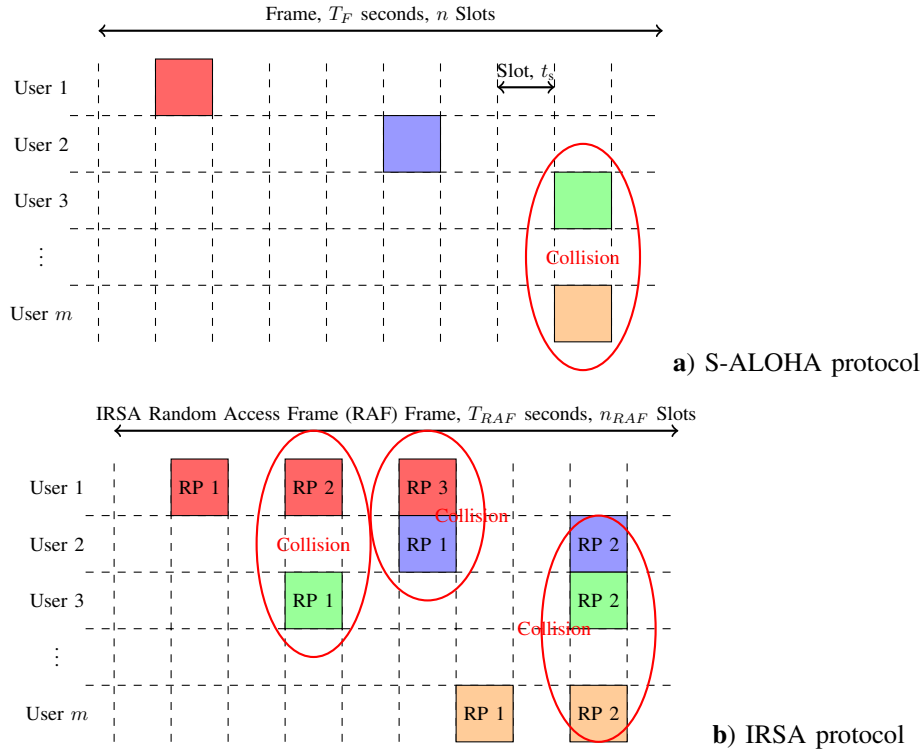

Fig. \ref{fig:S-ALOHA_IRSA} depicts a sketch of the slotted ALOHA and the \gls{IRSA} RA protocols. $n$ slots lasting $T_F$ seconds build a frame  $F$. The total number of active users is $m$. The MAC frame, also called Random Access Frame (RAF), of $T_F$ duration, is composed of $n$ slots of duration $t_\text{s}=T_F/n$. In every MAC frame, a measurable number of users $m$ try a packet transmission in a determined instant. Hence, in each MAC frame, \ta{each user} performs a unique transmission associated with a new packet or the retransmission of a previous collision. Further, a MAC frame with a collision does not allow retransmission. The last statement will generate some back-logged users.

The {\it normalized offered traffic} (or \ta{network} load) follows by:
\begin{equation} \label{eq:normG}
G=\frac{m}{n}  \qquad \left[\frac{\text{\ta{user}}}{\text{slot}}\right]
\end{equation}
Therefore, the {\it normalized throughput} (or channel output) $S$ is characterized as the probability of successful packet transmission per slot, where $n$ is the number of slots transmitted in $T_F$ seconds. Fig. \ref{fig:S-ALOHA_IRSA}.(a) presents an S-ALOHA frame with individual packet transmission by MAC frame. Collisions drive retransmissions in the subsequent frame. {Hence, the throughput for the S-ALOHA protocol can be defined} as a function of normalized offered traffic:
\begin{equation}
S(G)=G \cdot e^{-G}\qquad \ta{\left[\frac{\text{packet}}{\text{slot}}\right]}
\end{equation}
where the peak throughput is attained at $S(1)=e^{-1} \approx 0.37$.

\vspace{2mm}
\subsubsection{The Upbound of 0.37 for Slotted ALOHA}
{The upbound of 0.37 for slotted ALOHA has been extensively studied in the literature. Kennedy and Miller were the first to analyze the hidden terminal problem in carrier sense multiple-access and propose the busy-tone solution in 1975 \cite{Kennedy1975}. Daut and Shorey analyzed a random-access scheme for satellite networks and derived the upbound in 1981 \cite{Daut1981}. Eklund and Rahelić further studied the multi-access performance of packet radio networks with slotted ALOHA in 1986 \cite{Eklund1986}. More recently, Huang and van der Schaar proposed a compressive sensing-based approach for efficient and reliable random access in machine-to-machine communications in 2015 \cite{Huang2015}. Le and Poor developed a probabilistic method for medium access control in wireless networks and derived the upbound in 2019 \cite{Le2019}.  In recent years, several papers have focused on analyzing throughput for massive machine-type communications \cite{KimEtAl2020}, and applying mean-field game theory to distributed power control in wireless communication \cite{HassanEtAl2020}.}

\vspace{2mm}
\subsubsection{\gls{IRSA} Protocol} \label{sec:irsa}
The IRSA protocol relies on the repetition of each packet {access strategy} by MAC frame. In Fig. \ref{fig:S-ALOHA_IRSA}.b) each packet is transmitted $d$ times in a MAC frame. The repetition rate $d$ ranges from packet to packet based on a statistical distribution. Hence, we define the {\it user sampling degree distribution} by: 
\begin{equation}
\label{eq:degree_dist}
\Lambda(x)=\sum_{d=1}^{d_\text{m}}\Lambda_\text{d} x^d, \quad \text{with }
\quad \sum_{d=1}^{d_\text{m}}\Lambda_\text{d}=1
\end{equation}
where $0\leq \Lambda \leq 1$ 
{and each transmitter independently sends the replicas within the $n$ time slots composing the MAC frame. Random transmission slots contain a maximum of $d_\text{m}$ replicas sent. }

{Each packet contains a header with an index with the location of its copy. After receiving a packet successfully, the BS extracts the index identifying the replica positions. When packet replicas collide, they are extracted from the signal received in the corresponding slot removing the interference contribution. This procedure allows decoding packets transmitted in the same slot.}

\subsection{Traffic Model, Time Arrival, and Time Instance} \label{ssec:arrival}

\gls{MTC} traffic patterns differ from those for \gls{H2H} traffic. Better traffic models prompt better management of shared network resources and guarantee  \gls{QoS} for many types of devices. 3GPP proposes a simple Poisson process to model different kinds of network access coordinated or uncoordinated {machine-to-machine (M2M)} traffic \cite{Sunita2019}. Consequently,  {different} arrival rate $\lambda$ {dynamically depends on the channel and system} scenarios.   {5G system scenarios} are compiled in \cite{Navarro-Ortiz2020}.

3GPP Specification number 37.868 define \gls{M2M} traffic types, including a load analysis for \gls{SM}, fleet management, and earthquake monitoring applications. The document embraces a \gls{SM} service with metering data reports, in fixed time intervals, and the load control and alarm events occurring randomly, being modeled by Poisson processes. Table \ref{tab:traffic_models} compiles the two traffic models.

\begin{table}[htbp!]
	\centering
	\caption{3GPP traffic models for MTC proposed in the 3GPP TR 37.868 document.}
	\label{tab:traffic_models}
	\small
\begin{tabular}{ll}
\hline
\bf Parameter & \bf Statistical Characterization \\ \hline
			\multicolumn{2}{c}{\bf Traffic model 1} \\ \hline
			Number of MTC devices & ${M\in\, }\{1000,3000,5000,10000,30000\}$ \\
			Arrival distribution & Uniform distribution over $T_\text{A}=60$s \\
			$pk_s$ & 200 bytes \\ \hline
			\multicolumn{2}{c}{\bf Traffic model 2} \\ \hline
			Number of MTC devices & ${M\in\, }\{1000,3000,5000,10000,30000\}$ \\
			Arrival distribution & Beta distribution over $T_\text{A}=10$s \\
			$pk_s$ & 200 bytes
			\\ \hline
		\end{tabular}%
\end{table}

\begin{figure*}[!htbp]
    \centering
    \includegraphics[width=0.5\textwidth]{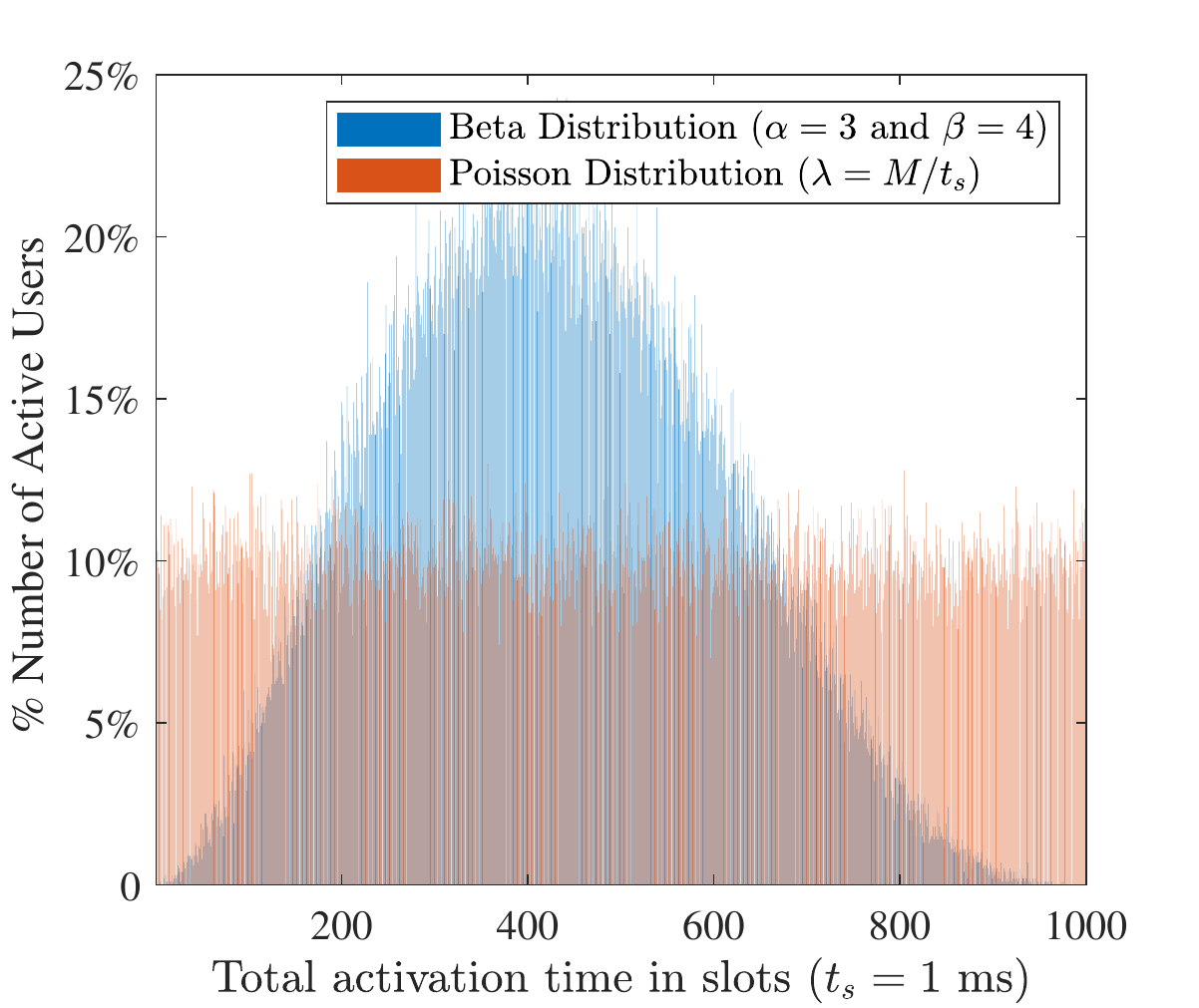} 
    \caption{{Users arrival in a Frame $F$ with $n$ time-slots considering Poisson $\times$ Beta distribution and $M=100$ devices.}}
    \label{fig:active_users}
\end{figure*}

Then \gls{IRSA} protocol design in \cite{Liva2011} considers Poisson traffic, {\it i.e.}, constant mean arrival even in \gls{SS} applications with a disrupted appearance. On the other hand,  the traffic model in \cite{Gursu2019} considers the Beta distribution that implicitly requires a time-varying mean parameter, \ta{{\it i.e.}, this traffic model accounts for the non-stationarity arrivals}. Hence, the arrival distribution of the $M_v$ active users in \gls{SG} applications can be modeled by a Poisson or Beta distributions \cite{Gursu2019,Gursu2019a} depending on the stationarity or  non-stationarity arrivals assumption. Typical but different behavior in terms of the number of user activation is found considering Poisson and Beta distribution, Fig. \ref{fig:active_users}.

After \ta{starting}, an IRSA frame does not permit seeking users. Each late-activated user must wait for a full-frame duration $T_F$. The existence of a waiting time $T_W$ within two IRSA frames relies on latency limitations.  Indeed the worst-case scenario implies a user decoded following the expiration of the IRSA frame.

Hereafter, the maximum latency $\ta{\Delta_i}$ of the user $i$ can be separated as
\begin{equation}
\ta{\Delta_i} = T_W + 2 T_F = T_\text{Ac} + T_F \leq  \ta{\Delta_0}
\end{equation}
A user's maximum latency should not exceed the latency constraint $\Delta_0$. {This scheme implies the accumulation of all users activated between \gls{IRSA} frames.}

\subsection{Successive Interference Cancellation (SIC)} \label{sec:SIC}

The scenarios involving M2M communications with smart sensors call for new ideas on RA schemes. The adoption of simple \gls{SIC} technique makes it feasible to deliver throughput enhancements in RA-based communication systems \cite{Mengali2018}. Based on the receiving and buffering MAC frame, the receiver looks for {\it singleton slots}, meaning slots with interference-free packets. 

\ta{In the context of RA protocols, a graphical model can be described as an  iterative \gls{SIC} process  \cite{Berioli2016}, where the MAC frame standing through a bipartite graph $\mathcal{G} = (\mathcal{U}, \mathcal{S}, \mathcal{E})$ composed by three sets:} 
\begin{itemize}
\item[$\mathcal{U}$]$= {\{U_1, U_2,\dots, U_m\}}$ of $m$ {\it user nodes} (UNs);
\item[$\mathcal{S}$]$ = {\{S_1, S_2,\dots, S_n\}}$ of $n$ {\it slot nodes} (SNs);
\item[$\mathcal{E}$] $ =$ {\it set of edges}.
\end{itemize}
%
The UN $U_i\in\mathcal{U}$ connects to the \gls{SN} $S_j\in\mathcal{S}$ by an edge $\mathcal{E}_{ij}$ conditioned to the copy of the {$i$th packet of the $i$th user is} transmitted in the $j$th slot. Fig. \ref{fig:bipartite} depicts the graph illustration of a MAC frame with $n = 5$ slots and $m = 4$ users attempting a transmission. {Notice that} there is a 2-gather for users 1, 2, and 4, while user 3 packets repeat three times. One {\it singleton slot} for the received frame. A particular slot node has degree $d$ if it is adjacent to $d$ edges{; for instance, in the example of} Fig. \ref{fig:bipartite}, {$S_1$} has degree {$d=2$}, and that $S_2$ has degree {$d=1$}.
\color{black}

\tikzset{square/.style={draw, regular polygon, regular polygon sides=4, minimum size=2cm}}
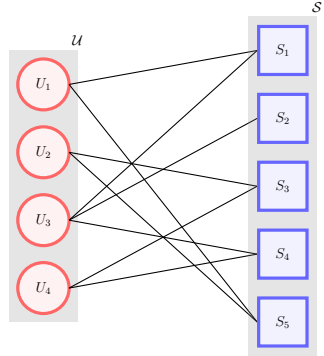
\begin{figure}[htbp]
\centering
\begin{tikzpicture}[scale=.45, transform shape]
        \fill[fill=gray, fill opacity=0.2] (1,1) rectangle (3,9) node[text=black, opacity=1,above] {$\mathcal{U}$};
        \filldraw[color=red!60, fill=red!5, very thick](2,2) circle (0.75) node[text=black, opacity=1] {$U_4$};
        \filldraw[color=red!60, fill=red!5, very thick](2,4) circle (0.75) node[text=black, opacity=1] {$U_3$};
        \filldraw[color=red!60, fill=red!5, very thick](2,6) circle (0.75) node[text=black, opacity=1] {$U_2$};
        \filldraw[color=red!60, fill=red!5, very thick](2,8) circle (0.75) node[text=black, opacity=1] {$U_1$};
        
        \draw [-] (2.75,8) to (8.25,9);
        \draw [-] (2.75,8) to (8.25,1);
        
        \draw [-] (2.75,6) to (8.25,5);
        \draw [-] (2.75,6) to (8.25,1);
        
        \draw [-] (2.75,4) to (8.25,9);
        \draw [-] (2.75,4) to (8.25,7);
        \draw [-] (2.75,4) to (8.25,3);
        
        \draw [-] (2.75,2) to (8.25,5);
        \draw [-] (2.75,2) to (8.25,3);
        
        \fill[fill=gray, fill opacity=0.2] (8,0) rectangle (10,10) node[text=black, opacity=1,above] {$\mathcal{S}$};

        \node[square,color=blue!60, fill=blue!5, very thick,text=black] at (9,1)  (0) {$S_5$};
        \node[square,color=blue!60, fill=blue!5, very thick,text=black] at (9,3) (0) {$S_4$};
        \node[square,color=blue!60, fill=blue!5, very thick,text=black] at (9,5) (0) {$S_3$};
        \node[square,color=blue!60, fill=blue!5, very thick,text=black] at (9,7) (0) {$S_2$};
        \node[square,color=blue!60, fill=blue!5, very thick,text=black] at (9,9) (0) {$S_1$};
    \end{tikzpicture}
    \caption{Bipartite graph for a MAC frame with {$n = 5$ slots and $m = 4$ users attempting} a transmission}
    \label{fig:bipartite}
\end{figure}

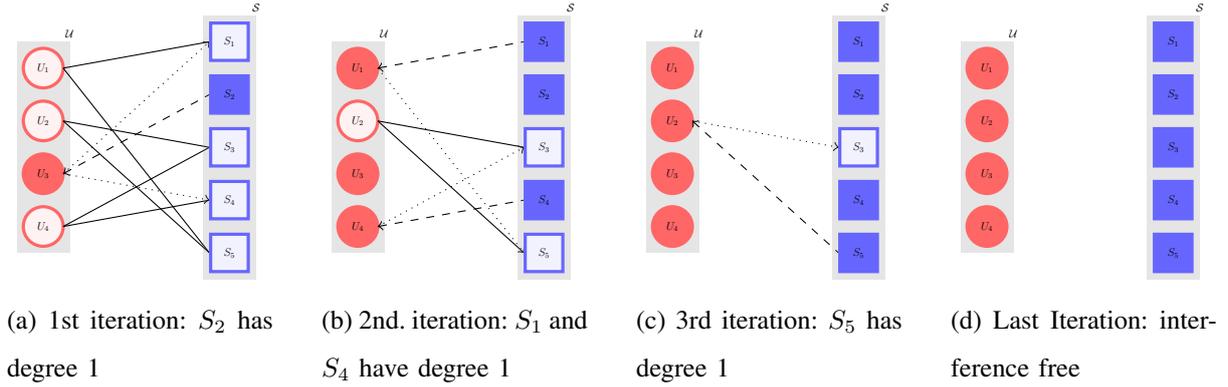
\begin{figure}[!htbp]
\centering
\begin{subfigure}[h]{.22\textwidth}
\centering
\begin{tikzpicture}[scale=.35, transform shape]
\color{black}
\fill[fill=gray, fill opacity=0.2] (1,1) rectangle (3,9) node[text=black, opacity=1,above] {$\mathcal{U}$};
\filldraw[color=red!60, fill=red!5, very thick](2,2) circle (0.75) node[text=black, opacity=1] {$U_4$};
\filldraw[color=red!60, fill=red!60, very thick](2,4) circle (0.75) node[text=black, opacity=1] {$U_3$};
\filldraw[color=red!60, fill=red!5, very thick](2,6) circle (0.75) node[text=black, opacity=1] {$U_2$};
\filldraw[color=red!60, fill=red!5, very thick](2,8) circle (0.75) node[text=black, opacity=1] {$U_1$};     
        \draw [-] (2.75,8) to (8.25,9);
        \draw [-] (2.75,8) to (8.25,1);
        \draw [-] (2.75,6) to (8.25,5);
        \draw [-] (2.75,6) to (8.25,1);
        \draw [dotted,->] (2.75,4) to (8.25,9);
        \draw [dashed,<-] (2.75,4) to (8.25,7);
        \draw [dotted,->] (2.75,4) to (8.25,3);
        \draw [-] (2.75,2) to (8.25,5);
        \draw [-] (2.75,2) to (8.25,3);
        \fill[fill=gray, fill opacity=0.2] (8,0) rectangle (10,10) node[text=black, opacity=1,above] {$\mathcal{S}$};
        \node[square,color=blue!60, fill=blue!5, very thick,text=black] at (9,1) (0) {$S_5$};
        \node[square,color=blue!60, fill=blue!5, very thick,text=black] at (9,3) (0) {$S_4$};
        \node[square,color=blue!60, fill=blue!5, very thick,text=black] at (9,5) (0) {$S_3$};
        \node[square,color=blue!60, fill=blue!60, very thick,text=black] at (9,7) (0) {$S_2$};
        \node[square,color=blue!60, fill=blue!5, very thick,text=black] at (9,9) (0) {$S_1$};
\end{tikzpicture}
\caption{1st iteration:  $S_2$ has degree 1}
\label{fig:SIC_a}
\end{subfigure}
\hspace{3.5mm}
\begin{subfigure}[h]{0.22\textwidth}
\centering
\begin{tikzpicture}[scale=.35, transform shape]
\color{black}
\fill[fill=gray, fill opacity=0.2] (1,1) rectangle (3,9) node[text=black, opacity=1,above] {$\mathcal{U}$};
\filldraw[color=red!60, fill=red!60, very thick](2,2) circle (0.75) node[text=black, opacity=1] {$U_4$};
\filldraw[color=red!60, fill=red!60, very thick](2,4) circle (0.75) node[text=black, opacity=1] {$U_3$};
        \filldraw[color=red!60, fill=red!5, very thick](2,6) circle (0.75) node[text=black, opacity=1] {$U_2$};
        \filldraw[color=red!60, fill=red!60, very thick](2,8) circle (0.75) node[text=black, opacity=1] {$U_1$};
        
        \draw [dashed,<-] (2.75,8) to (8.25,9);
        \draw [dotted,->] (2.75,8) to (8.25,1);
        
        \draw [-] (2.75,6) to (8.25,5);
        \draw [-] (2.75,6) to (8.25,1);

        \draw [dotted,->] (2.75,2) to (8.25,5);
        \draw [dashed,<-] (2.75,2) to (8.25,3);

\fill[fill=gray, fill opacity=0.2] (8,0) rectangle (10,10) node[text=black, opacity=1,above] {$\mathcal{S}$};
        \node[square,color=blue!60, fill=blue!5, very thick,text=black] at (9,1) (0) {$S_5$};
        \node[square,color=blue!60, fill=blue!60, very thick,text=black] at (9,3) (0) {$S_4$};
        \node[square,color=blue!60, fill=blue!5, very thick,text=black] at (9,5) (0) {$S_3$};
        \node[square,color=blue!60, fill=blue!60, very thick,text=black] at (9,7) (0) {$S_2$};
        \node[square,color=blue!60, fill=blue!60, very thick,text=black] at (9,9) (0) {$S_1$};
    \end{tikzpicture}
\caption{2nd. iteration:  $S_1$ and $S_4$ have degree 1}
\label{fig:SIC_b}
\end{subfigure}
\hspace{3.5mm} 
\begin{subfigure}[h]{0.22\textwidth}
\centering
\begin{tikzpicture}[scale=.35, transform shape]
\color{black}
\fill[fill=gray, fill opacity=0.2] (1,1) rectangle (3,9) node[text=black, opacity=1,above] {$\mathcal{U}$};
        \filldraw[color=red!60, fill=red!60, very thick](2,2) circle (0.75) node[text=black, opacity=1] {$U_4$};
        \filldraw[color=red!60, fill=red!60, very thick](2,4) circle (0.75) node[text=black, opacity=1] {$U_3$};
        \filldraw[color=red!60, fill=red!60, very thick](2,6) circle (0.75) node[text=black, opacity=1] {$U_2$};
        \filldraw[color=red!60, fill=red!60, very thick](2,8) circle (0.75) node[text=black, opacity=1] {$U_1$};
        \draw [dotted,->] (2.75,6) to (8.25,5);
        \draw [dashed,<-] (2.75,6) to (8.25,1);
        \fill[fill=gray, fill opacity=0.2] (8,0) rectangle (10,10) node[text=black, opacity=1,above] {$\mathcal{S}$};
        \node[square,color=blue!60, fill=blue!60, very thick,text=black] at (9,1) (0) {$S_5$};
        \node[square,color=blue!60, fill=blue!60, very thick,text=black] at (9,3) (0) {$S_4$};
        \node[square,color=blue!60, fill=blue!5, very thick,text=black] at (9,5) (0) {$S_3$};
        \node[square,color=blue!60, fill=blue!60, very thick,text=black] at (9,7) (0) {$S_2$};
        \node[square,color=blue!60, fill=blue!60, very thick,text=black] at (9,9) (0) {$S_1$};
    \end{tikzpicture}
         \caption{3rd iteration: $S_5$ has degree 1}
         \label{fig:SIC_c}
     \end{subfigure}
\hspace{3.5mm}  
\begin{subfigure}[h]{0.22\textwidth}
\centering
\begin{tikzpicture}[scale=.35, transform shape]
\color{black}
 \fill[fill=gray, fill opacity=0.2] (1,1) rectangle (3,9) node[text=black, opacity=1,above] {$\mathcal{U}$};
        \filldraw[color=red!60, fill=red!60, very thick](2,2) circle (0.75) node[text=black, opacity=1] {$U_4$};
        \filldraw[color=red!60, fill=red!60, very thick](2,4) circle (0.75) node[text=black, opacity=1] {$U_3$};
        \filldraw[color=red!60, fill=red!60, very thick](2,6) circle (0.75) node[text=black, opacity=1] {$U_2$};
        \filldraw[color=red!60, fill=red!60, very thick](2,8) circle (0.75) node[text=black, opacity=1] {$U_1$};

        \fill[fill=gray, fill opacity=0.2] (8,0) rectangle (10,10) node[text=black, opacity=1,above] {$\mathcal{S}$};
        \node[square,color=blue!60, fill=blue!60, very thick,text=black] at (9,1) (0) {$S_5$};
        \node[square,color=blue!60, fill=blue!60, very thick,text=black] at (9,3) (0) {$S_4$};
        \node[square,color=blue!60, fill=blue!60, very thick,text=black] at (9,5) (0) {$S_3$};
        \node[square,color=blue!60, fill=blue!60, very thick,text=black] at (9,7) (0) {$S_2$};
        \node[square,color=blue!60, fill=blue!60, very thick,text=black] at (9,9) (0) {$S_1$};
\end{tikzpicture}
\caption{Last Iteration: interference free} \label{fig:SIC_d}
\end{subfigure}
\caption{SIC decoding for the collision pattern of Fig. \ref{fig:bipartite}, tracked over the corresponding graph\ta{; four SIC iterations are represented}.}
\label{fig:SIC_dec}
\end{figure}

To model the \gls{SIC} process through a graph, some simple rules must be applied.  Iteratively, we search for \gls{SN}s with degree $d=1$. Estimate $S_i$ to have degree 1, and indicate by $U_j$ its only \ta{user node} neighbor.  Hence, we exclude the edges connected to  $S_i$ and $U_j$; equal to exclude the interference addition on the packet of user $j$ in each transmitted slot. {The \gls{SIC} process} repeats on the up-to-date graph. \ta{Following these rules, the \gls{SIC} process applied to the example of Fig. \ref{fig:bipartite} is sketched as a four canceling iterations process represented as graph model in Fig. \ref{fig:SIC_dec}.}

The fundamental concept of \gls{SIC} is decoding {users' signal (devices)} successively. After decoding one {user}, its signal is reconstructed and canceled from the total received signal before decoding the subsequent {user's signal}.  In this work, we use the advantage of a singleton where at least one device is free of interference. Hence, assuming conventional \gls{SIC} application, one of the users \ta{(strongest sinal)}, say \ta{$S_1$}, is decoded treating \ta{$S_2, S_3,\ldots, S_n$ as interference. Next, the second strongest user' signal,} \ta{$S_2$}, is decoded by using the \ta{extracted, reconstructed, and subtracted \ta{$S_1$} signal from the total received signal. The canceling process is repeated until the last interfering signal is decoded}.  SIC implementation represents an advantage compared to the conventional decoding process, which treats all the interfering users' signals as noise.

\subsection{Time-domain Structure}
The parameters of \gls{SG} system considered herein are based on the 5G \gls{NR} standard \cite{Dahlman2018}, \cite{NR_PHY-3GPP}.  Time-domain NR transmissions take place into $10$ ms length frames. {The radio frame of 10 ms has transmitted continuously as per TDD topology.} Each frame contains $10$ equally sized subframes of duration $1$ ms.  
\gls{ScS} establishes the bandwidth of each frame  ($BW_\text{F}$) and the number of slots in a subframe. With slot-based scheduling, a slot is a minimum measure with a duration of $t_\text{s}$ seconds.
{A subframe is of fixed duration {\it i.e.} 1ms, whereas {\it slot length} varies based on subcarrier spacing and the number of slots per subframe.  It is 1 ms for 15 KHz, 500 $\mu$s for 30 KHz, and so on. Subcarrier spacing of 15 KHz occupies 1 slot per subframe, subcarrier spacing of 30 KHz occupy 2 slots per subframe, and so on. }
Finally, this work does not use a symbol-level scheduling technique.

\section{Metrics and Definitions in \gls{SG} \gls{QoS}}\label{sec:QoS_Metrics}
\subsection{Latency}
From the power system perspective, the latency requirement depends on the cycle of the electrical utility cycle ($T = 1/f $) to keep the stable stage of a wave. 
\ta{Generically, in the context of a smart grid, we define data (or message) latency as the time between a state's occurrence and an application's activation \cite{Kansal2012}. Each
application has its own latency requirements depending upon
the kind of system response it is dealing with. Besides, among the other delays, {\it communication delay} also adds to the latency and needs to be minimized}.

The total delay on the communication network comprises propagation delays, transmission delays, queuing delays, and processing delays. 
Primarily, {\it latency} ($L$) represented by:
\begin{equation} \label{eq:latency}
     L =\tau \, \frac{1}{f} 
\end{equation}
{where $\tau$ is the delay factor (in cycles).  Real-time scale \gls{SG} network considers short/small values of $\tau$.}

{We define in the following the \gls{QoS} metrics RA-related to \gls{SG} applications: the latency of message,  \gls{PDR} reliability, and critical latency response.}

\subsubsection{{Latency in SG Message}}
This work focuses on applications with communication latency performance as a priority and does not accept outdated data. Thus, we define the hard delay metric in the subsequent outlines.

\begin{defn} \label{def:HD}
\textbf{(Hard delay, HD)} {Application with a hard delay requirement turns useless if its delay passes the QoS requirement even in successful message delivery.}
\end{defn}

{Next, the delay requirement $\tau_{\rm req}$ includes the required network bandwidth as:}
\begin{equation}\label{eq:BWreq}
    BW_{\rm req}=S \, \left(8\frac{\rm bits}{\rm byte}\right) \, \frac{1}{\tau_{\rm req}} \, M \qquad \rm {[bits\cdot Hz]}
\end{equation}
{where $S$ and $M$ are the service's data size in bytes, and the number of users, respectively.}

\subsection{Packet Delivery \ta{Ratio} (PDR) and Packet Loss \ta{Ratio} (PLR)} 

\begin{defn} \label{def:PDR}
\textbf{Packet Delivery \ta{Ratio} (PDR)} represents the proportion of the total number of packets received by the \gls{DCU} successfully, $P_{\rm R}$, to the total number of packets generated by the \gls{SG} source nodes, {$P_G$}:
\begin{equation} \label{eq:PDR}
PDR=\frac{\sum^{M}_{i=1}P_{\text{R}_i}}{\sum^{M}_{i=1}P_{G_i}}=\frac{P_\text{R}}{P_\text{G}}
\end{equation}
\end{defn}

\begin{defn} \label{def:PLR}
\textbf{\gls{PLR}} {is a matched metric to the PDR defined as the probability for a user incorrectly decoded at the receiver after accessed the channel.} Using eq. \eqref{eq:PDR} one can obtain an approximate expression for PLR:
\begin{equation} \label{eq:PLR}
PLR=1-PDR=1-\frac{P_\text{R}}{P_\text{G}}
\end{equation}
\end{defn}

\subsection{Reliability} 

A conventional reliability definition in computer systems deliver \gls{SG} network reliability metric \cite{Sahner1996}. Network reliability counts the probability of a system achieving its services accordingly in a specified time duration. Accordingly, we define the {\it reliability factor} in the \gls{SG} network context as:
\begin{defn} \label{def:R}
\textbf{(Reliability, {$\mathcal{R}$})} {The network reliability stands as the probability of succeeding in delivering a message to the destinations node ({related to the} PDR metric) within the \ta{maximum latency} requirement. The reliability decreases when the message arrives after the delay requirement.}
\gls{SG} network reliability {under} HD {definition}, $\tau_{\text{HD}_{\rm req}}$ can be defined as follows. 

\begin{equation}\label{eq:reliability_1}
   {\mathcal{R}(\tau)} =\left\{
    \begin{array}{ll}
    PDR \,\, \ta{ = 1-PLR}& \tau \leq \tau_{\text{HD}_{\rm req}} \\
    0 & \tau > \tau_{\text{HD}_{\rm req}}
    \end{array}
    \right.
\end{equation}
\end{defn}
\ta{Notice that in {\it ultra-reliable and low latency communications} (URLLC), a generic reliability requirement of $\mathcal{R} = 1-10^{-5}$ ({\it i.e.}, 0.99999) is required with the user-plane radio-latency of 1 ms for a single transmission of a 32-byte long packet}.

\begin{defn} \label{def:ACR}
\textbf{\gls{ACR}}. We introduce a new concept named \gls{ACR} that represents the number of users to  successfully comply with the \gls{UL} connection within delay requirements $\tau_{\text{HD}_{\rm req}}$. In the context of SG applications, the \gls{ACR} reliability can be defined \ta{by the sum of ratios along the $N$ slots}:
\begin{equation} \label{eq:ACR}
\mathcal{R}_{\text{ACR}} = \frac{1}{N}\sum_{k=1}^{N}\frac{\ta{m(\tau_k)}}{m_k}
\end{equation}
\ta{where $m_k$ is the number of total active users in the $k$-th time-slot, and
$m\ta{(\tau_k)}$ represents the number of users that effectively sent a packet at the $k$th time-slot $k\in\{1,\ldots, N\}$ within the latency requirement time, {\it i.e.},  $\tau_k\leq \tau_{{\rm req}_i}$.} Such reliability metric quantifies the proportion of  active users who succeed in sending a packet at the $k$th time-slot 
within the latency requirement time \ta{along $N$ slots.}
\end{defn}
In eq. (\ref{eq:ACR}), the \gls{QoS} metric \ta{allows to track specific application problems.  Under specific protocol applications, the ACR reliability metric quantifies the trusted network data} delivery.

\section{Proposed RA Protocol \ta{based on Raptor Codes} for \gls{m-SGC} Networks} \label{sec:Proposed_RA}
In this section, we propose a new RA protocol to achieve higher throughput {under} overloaded networks using additional nodes working as precoding of Raptor codes. This Raptor code implementation uses a subclass of the bipartite graph described in \cite{Ramatryana2017, Shokrollahi2006}.

{Considering} \gls{SG} scenarios, {in the sequel we describe briefly the Raptor codes, including their characteristics, advantages, and some application environments with the associated parameters. Then,} we propose a modification in the traditional \gls{IRSA} protocol, mainly in the values of the operational parameters and in the substructure of the protocol,  depending on the application requirements,  aiming at enhancing the protocol reliability suitable for {\it ultra-reliable low-latency communications} (URLLC) and {\it machine-type communications} (MTC) services.  We label \gls{QoS} information into active users preamble. This information adjusts the \gls{IRSA} protocol to achieve \gls{QoS} requirements.

\subsection{{Raptor Codes}}\label{sec:Raptor} 
{Raptor codes (RapC) are a class of forward error correction codes that can provide reliable and efficient data transmission over unreliable communication channels \cite{WangEtAl2019}. These codes were introduced by A. Shokrollahi in 2003 \cite{Shokrollahi2006}. RapC can be used to efficiently encode and decode data packets for transmission over noisy channels. They are based on the concept of fountain codes and use an iterative decoding process to recover the original data. RapC's {\it advantages} over traditional error correction codes include bandwidth efficiency, higher reliability, and less decoding complexity. RapC can also adapt to different channel conditions and can handle varying levels of noise and interference. Raptor codes have been widely used in various communication systems, including wireless networks, multimedia transmission, and satellite communications. In 5G mobile networks, Raptor codes are used in the control plane and user plane to improve reliability and reduce latency \cite{3GPP.38.214}.}

{The key parameters involved in RapC include the degree distribution, symbol size, and encoding rate. The {\it degree distribution} determines the distribution of probability for the number of times each symbol appears in the code word. The {\it symbol size} refers to the size of the symbols in the code word. The {\it encoding rate} determines the ratio of the output code length to the input message length. The encoding process of Raptor codes is based on a Fountain code, which generates a potentially infinite number of encoded packets from a source packet \cite{ChenEtAl2020}. A decoder can recover the original message from a subset of these encoded packets. The {\it repetition rate} of Raptor code, {\it i.e.}, the number of times a message is encoded into the code is denoted by $\eta$. Raptor codes have been used in various application environments, including satellite and wireless networks \cite{XuEtAl2018,LiEtAl2019}. In the RapIRSA protocol, the {\it number of connecting nodes} ($c\mathcal{N}$) used is denoted by $q$ \cite{DattaEtAl2020}.}

{Overall, Raptor codes are a powerful tool for providing reliable and efficient data transmission in various communication systems and channel scenarios. Their adaptability, high reliability, and low decoding complexity make them an attractive option for 5G mobile networks and beyond.}

\subsection{{Combining} Raptor Codes and {IRSA} (RapIRSA)} \label{sec:RapC}

We consider $m$ active users sending packets to BS or DCU. Some users will send packets via {\it connecting nodes} ($c\mathcal{N}$). All $c\mathcal{N}$ are placed equitably reaching many possible directly visible users. Each $c\mathcal{N}$ uses the Decode-and-Forward (DF) protocol to deliver the information to the BS. 
$c\mathcal{N}$ has less probability of packet collisions compared to BS because $c\mathcal{N}$ only received packets of data from connected users.

\begin{figure}[!htbp]
\centering
\begin{tikzpicture}[scale=.4, transform shape]
\color{black}
\filldraw[color=red!60, fill=red!5, very thick](-5,1) circle (0.75) node[text=black, opacity=1] {$U_5$};
\filldraw[color=red!60, fill=red!5, very thick](-5,3) circle (0.75) node[text=black, opacity=1] {$U_4$};
\filldraw[color=red!60, fill=red!5, very thick](-5,5) circle (0.75) node[text=black, opacity=1] {$U_3$};
\filldraw[color=red!60, fill=red!5, very thick](-5,7) circle (0.75) node[text=black, opacity=1] {$U_2$};
\filldraw[color=red!60, fill=red!5, very thick](-5,9) circle (0.75) node[text=black, opacity=1] {$U_1$};
        \draw [-] (-4.25,1) to (1.25,1);
        \draw [-] (-4.25,3) to (1.25,3);
        \draw [-] (-4.25,5) to (1.25,5);
        \draw [-] (-4.25,7) to (1.25,7);
        \draw [-] (-4.25,9) to (1.25,9);
\fill[fill=gray, fill opacity=0.2] (1,0) rectangle (3,10) node[text=black, opacity=1,above] {$\mathcal{U}$};
\filldraw[color=red!60, fill=red!5, very thick](2,1) circle (0.75) node[text=black, opacity=1] {$U_5$};
\filldraw[color=red!60, fill=red!5, very thick](2,3) circle (0.75) node[text=black, opacity=1] {$U_4$};
\filldraw[color=red!60, fill=red!5, very thick](2,5) circle (0.75) node[text=black, opacity=1] {$U_3$};
\filldraw[color=red!60, fill=red!5, very thick](2,7) circle (0.75) node[text=black, opacity=1] {$U_2$};
\filldraw[color=red!60, fill=red!5, very thick](2,9) circle (0.75) node[text=black, opacity=1] {$U_1$};    
        \draw [-] (2.75,9) to (8.25,9);
        \draw [-] (2.75,9) to (8.25,7);
        \draw [-] (2.75,9) to (8.25,5);
        \draw [-] (2.75,7) to (8.25,9);
        \draw [-] (2.75,7) to (8.25,7);
        \draw [-] (2.75,7) to (8.25,5);
        \draw [-] (2.75,5) to (8.25,9);
        \draw [-] (2.75,5) to (8.25,3);
        \draw [dotted,-] (2.75,5) to (8.25,1);
        \draw [-] (2.75,3) to (8.25,7);
        \draw [-] (2.75,3) to (8.25,3);
        \draw [dotted,-] (2.75,3) to (8.25,1);
        \draw [-] (2.75,1) to (8.25,5);
        \draw [dotted,-] (2.75,1) to (8.25,1);
        \fill[fill=gray, fill opacity=0.2] (8,0) rectangle (10,10) node[text=black, opacity=1,above] {$\mathcal{S}$};
        \node[square,color=blue!60, fill=blue!5, very thick,text=black] at (9,1) (0) {$S_5$};
        \node[square,color=blue!60, fill=blue!5, very thick,text=black] at (9,3) (0) {$S_4$};
        \node[square,color=blue!60, fill=blue!5, very thick,text=black] at (9,5) (0) {$S_3$};
        \node[square,color=blue!60, fill=blue!5, very thick,text=black] at (9,7) (0) {$S_2$};
        \node[square,color=blue!60, fill=blue!5, very thick,text=black] at (9,9) (0) {$S_1$};
        
        \draw [-] (-4.25,9) to (1.25,-4);
        \draw [-] (-4.25,7) to (1.25,-2);
        \draw [-] (-4.255,5) to (1.25,-2);
        \draw [-] (-4.255,1) to (1.25,-4);
        
        \fill[fill=gray, fill opacity=0.2] (1,-5) rectangle (3,-1) node[text=black, opacity=1,above] {$c\mathcal{N}$};
        \filldraw[color=black!60, fill=gray!50, very thick](2,-2) circle (0.75) node[text=black, opacity=1] {$c\mathcal{N}_2$};
        \filldraw[color=black!60, fill=gray!50, very thick](2,-4) circle (0.75) node[text=black, opacity=1] {$c\mathcal{N}_1$};    

        \draw [-] (2.75,-2) -- (8.25,-2) node[pos=0.5,above] {$n_2$} ;
        \draw [-] (2.75,-4) -- (8.25,-4) node[pos=0.5,above] {$n_1$} ;
        \fill[fill=gray, fill opacity=0.2] (8,-5) rectangle (10,-1) node[text=black, opacity=1,above] {$\mathcal{S_{\text{c}\mathcal{N}}}$};
        \node[square,color=black!60, fill=blue!5, very thick,text=black] at (9,-2) (0) {$S_{c\mathcal{N}_2}$};
        \node[square,color=black!60, fill=blue!5, very thick,text=black] at (9,-4) (0) {$S_{c\mathcal{N}_1}$};
    \end{tikzpicture}
\caption{Raptor codes-structured wireless networks. \ta{RapIRSA$(\eta,q)$ parameters: $\eta= \frac{1}{7}$ and $q=2$}.}
\label{fig:Raptor}
\end{figure}
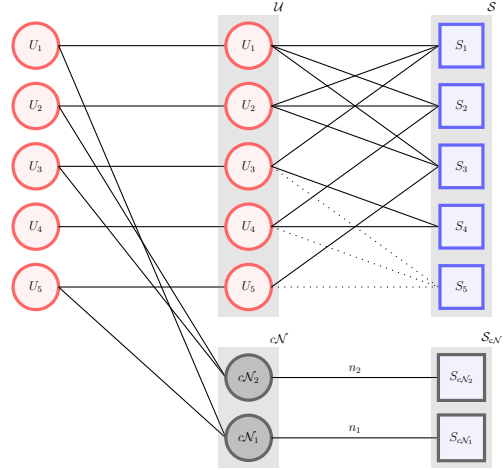

{Fig.} \ref{fig:Raptor} illustrates the \gls{RapC} wireless networks. This follows a bipartite graph model illustrated in {Fig.} \ref{fig:bipartite}. Besides, $c\mathcal{N}$ represented by dark circles in  a separated time slot;  $m$ {represents} the number of active users, $n_{\rm RapC}$ is \ta{the number of slot nodes (\gls{SN})} in the frame, {and} {$q=\floor{\eta \cdot n_{\rm RAF}}$} is the number of $c\mathcal{N}$. {Besides,} $n_{\rm RapC}$ contains two elements, as $n_{\textsc{raf}}$ for \gls{SN} of \gls{RAF}, and $n_q$ for \gls{SN} of $c\mathcal{N}$s.
{Hence,} the total \gls{SN} is defined as:
\begin{align}
    n_{\rm RapC} &= n_{\textsc{raf}} +n_q
\end{align}
\ta{Besides,} the normalized offered traffic $G$ \ta{for the RapIRSA follows the definition in Eq.} \eqref{eq:normG}.

{In the example of {Fig.} \ref{fig:Raptor},} five users $\mathcal{U}={\{U_1, U_2,\ldots,{U_5}\}}$ and two $c\mathcal{N}=\{c\mathcal{N}_1,c\mathcal{N}_2\}$ {have been considered}. Connecting nodes $c\mathcal{N}$ do not have prior information about users' messages. After receiving data from neighboring devices they {decode-and-forward (DF) the} messages to the BS for final processing. $c\mathcal{N}_1$ decodes one packet from $U_1$ and $U_5$, and then forward the first decoded packet to the BS. Likewise, $c\mathcal{N}_2$ receives packets from $U_2$ and $U_3$, {it decodes} one packet, and then forwards the decoded packets to BS. Fig. \ref{fig:Raptor} {represents} RapC {with} $\eta= \frac{1}{7}$ and $q=2$\ta{, {\it i.e.,} the parameters RapIRSA$(\eta,q)$ stand for the fraction of the number of IRSA random access frame\footnote{Namely repetition rate of Raptor code.} (RAF) length and the number of connecting nodes ($c\mathcal{N}$), respectively.}

{In the Raptor codes-based RapIRSA scheme, $c\mathcal{N}_1$ and $c\mathcal{N}_2$ do not have prior information about the messages transmitted by users. They receive packets from neighboring devices and decode and forward (DF) them to the BS for final processing. In the case of collisions, the RapC enable them to decode as much information as possible from the received packets, even if the packets are partially corrupted due to collisions. Moreover, efficient decoding algorithms such as belief propagation or message passing can be used at the connecting nodes to mitigate the effect of collisions and improve decoding performance. Recent research has shown the effectiveness of such algorithms in dealing with collisions in various wireless communication scenarios \cite{Wang2021, Zhang2021}.}

\subsubsection{Practicability}
{The proposed RapIRSA scheme requires each random access node to transmit its packet to three nodes (BS, $c\mathcal{N}_1$, and $c\mathcal{N}_2$). However, we believe that the proposed method is practical in real-world wireless network scenarios. The transmission scheme used in RapIRSA is similar to that used in other existing schemes such as CRDSA, which has been shown to be practical and effective in different network scenarios \cite{Beytur2020, Niu2012}. Furthermore, RapC used in our proposed scheme has been shown to improve reliability and reduce overhead in several other contexts of application \cite{Shokrollahi2006, sridharan2008}. Moreover, our proposed RapIRSA scheme is designed to be scalable and flexible to accommodate different numbers of users and connecting nodes. Efficient algorithms such as {\it belief propagation} or {\it message passing} can be used for the decoding process at the connecting nodes \cite{Richardson2008}. Therefore, we believe that our proposed RapIRSA scheme is practical and can be implemented in various real-world scenarios.}

\begin{figure}[!htbp]
\small
\centering
\begin{tikzpicture}[scale=.7, transform shape]
       \color{black}
        \draw[dashed,step=1cm,black,thin] (3,-3) grid (13.1,2.99);
        \path (2,2.5) node(u1) {User 1};
        \path (2,1.6) node(u) {\textbf{$\boldsymbol{\vdots}$}};
        \path (2,0.5) node(um) {User $m$};
         \draw[-,black] (2,0) -- (13.1,0) ;
        \path (2,-0.5) node(cN) {$c\mathcal{N}$ 1};
        \path (2,-1.4) node(c) {\textbf{$\boldsymbol{\vdots}$}};
        \path (2,-2.5) node(cN) {$c\mathcal{N}$ $q$};
        \draw[thick,<->,black] (3,3.5) -- (9.9,3.5) node[pos=0.5,above] {$T_{RAF}$ [s], $n_{RAF}$ [Slots]};
        \draw[thick,<->,black] (10,3.5) -- (12.9,3.5) node[pos=0.5,above] {$T_q$ [s], $n_q$ [Slots]};
        \filldraw[fill=blue!40!white, draw=black] (3,2) rectangle (4,3) node[pos=0.5] {RP 1};
        \filldraw[fill=blue!40!white, draw=black] (8,2) rectangle (9,3) node[pos=0.5] {RP $d_1$};
        \filldraw[fill=blue!40!white, draw=black] (4,0) rectangle (5,1) node[pos=0.5] {RP 1};
        \filldraw[fill=blue!40!white, draw=black] (8,0) rectangle (9,1) node[pos=0.5] {RP $d_\text{m}$};
        
        \path (6.5,2.5) node(c) {\textbf{$\boldsymbol{\dots}$}};
        \path (6.5,1.5) node(c) {\textbf{$\boldsymbol{\ddots}$}};
        \path (6.5,0.5) node(c) {\textbf{$\boldsymbol{\dots}$}};
        \path (6.5,-0.5) node(c) {\textbf{$\boldsymbol{\dots}$}};
        \path (6.5,-1.5) node(c) {\textbf{$\boldsymbol{\ddots}$}};
        \path (6.5,-2.5) node(c) {\textbf{$\boldsymbol{\dots}$}};

        \draw[-,black] (10,-3) -- (10,3) ;
        \filldraw[fill=gray!40!white, draw=black] (10,-1) rectangle (11,0) node[pos=0.5] {$S_{c \mathcal{N}_1}$};
        \path (11.5,-1.4) node(c) {\textbf{$\boldsymbol{\ddots}$}};
        \path (11.5,1.5) node(c) {\textbf{$\boldsymbol{\dots}$}};
        \filldraw[fill=gray!40!white, draw=black] (12,-3) rectangle (13,-2) node[pos=0.5] { $S_{c \mathcal{N}_q}$};
    \end{tikzpicture}
    \caption{Overview of \gls{RapIRSA} time-slots structure.}  
    \label{fig:RapIRSA_time}
\end{figure}
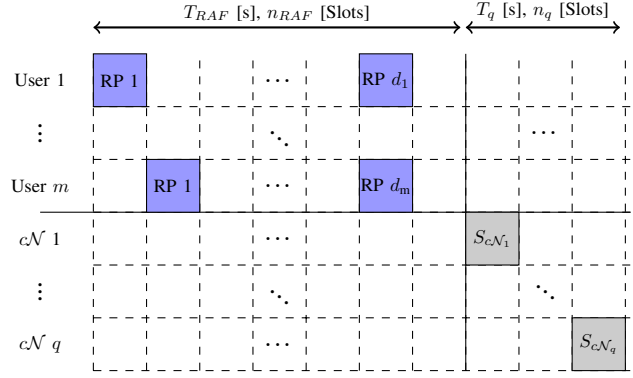

{Fig.} \ref{fig:RapIRSA_time} {depicts} the time-slot structure including the \gls{RAF} implemented in the \gls{IRSA} protocol and the slots with information from the $c\mathcal{N}$s. BS uses  pre-coding through $c\mathcal{N}$s to {detect} failing nodes during the contention period. {As discussed in the {numerical results}, Section \ref{sec:results}, the} \gls{RapIRSA} {protocol is able to improve} throughput performance substantially under high network load configurations, \ta{\textit{i.e.}, $G>0.6$}.

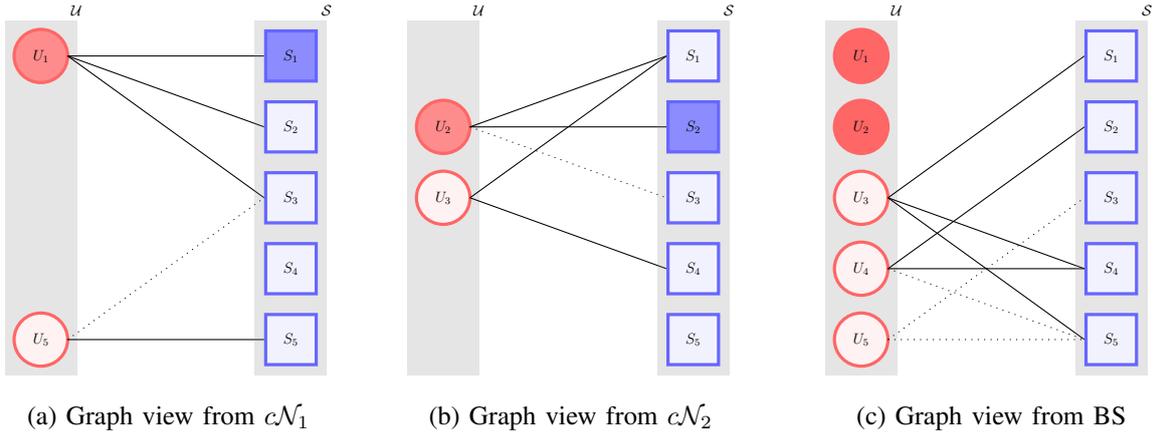
\begin{figure}[!htbp]
\centering
\begin{subfigure}[h]{.27\textwidth}
\centering
\begin{tikzpicture}[scale=.47, transform shape]
\color{black}
\fill[fill=gray, fill opacity=0.2] (1,0) rectangle (3,10) node[text=black, opacity=1,above] {$\mathcal{U}$};
\filldraw[color=red!60, fill=red!5, very thick](2,1) circle (0.75) node[text=black, opacity=1] {$U_5$};
\filldraw[color=red!60, fill=red!45, very thick](2,9) circle (0.75) node[text=black, opacity=1] {$U_1$};    
            \draw [-] (2.75,9) to (8.25,9);
            \draw [-] (2.75,9) to (8.25,7);
            \draw [-] (2.75,9) to (8.25,5);
            \draw [dotted,-] (2.75,1) to (8.25,5);
            \draw [-] (2.75,1) to (8.25,1);
            \fill[fill=gray, fill opacity=0.2] (8,0) rectangle (10,10) node[text=black, opacity=1,above] {$\mathcal{S}$};
            \node[square,color=blue!60, fill=blue!5, very thick,text=black] at (9,1) (0) {$S_5$};
            \node[square,color=blue!60, fill=blue!5, very thick,text=black] at (9,3) (0) {$S_4$};
            \node[square,color=blue!60, fill=blue!5, very thick,text=black] at (9,5) (0) {$S_3$};
            \node[square,color=blue!60, fill=blue!5, very thick,text=black] at (9,7) (0) {$S_2$};
            \node[square,color=blue!60, fill=blue!45, very thick,text=black] at (9,9) (0) {$S_1$};
\end{tikzpicture}
\caption{Graph view from $c\mathcal{N}_1$}
\label{fig:Raptor_cN1}
\end{subfigure}
 \hspace{7mm}
\begin{subfigure}[h]{.27\textwidth}
\centering 
\begin{tikzpicture}[scale=.47, transform shape]
            \color{black}
            \fill[fill=gray, fill opacity=0.2] (1,0) rectangle (3,10) node[text=black, opacity=1,above] {$\mathcal{U}$};
            \filldraw[color=red!60, fill=red!5, very thick](2,5) circle (0.75) node[text=black, opacity=1] {$U_3$};
            \filldraw[color=red!60, fill=red!45, very thick](2,7) circle (0.75) node[text=black, opacity=1] {$U_2$};
            \draw [-] (2.75,7) to (8.25,9);
            \draw [-] (2.75,7) to (8.25,7);
            \draw [dotted,-] (2.75,7) to (8.25,5);
            \draw [-] (2.75,5) to (8.25,9);
            \draw [-] (2.75,5) to (8.25,3);
            \fill[fill=gray, fill opacity=0.2] (8,0) rectangle (10,10) node[text=black, opacity=1,above] {$\mathcal{S}$};
            \node[square,color=blue!60, fill=blue!5, very thick,text=black] at (9,1) (0) {$S_5$};
            \node[square,color=blue!60, fill=blue!5, very thick,text=black] at (9,3) (0) {$S_4$};
            \node[square,color=blue!60, fill=blue!5, very thick,text=black] at (9,5) (0) {$S_3$};
            \node[square,color=blue!60, fill=blue!45, very thick,text=black] at (9,7) (0) {$S_2$};
            \node[square,color=blue!60, fill=blue!5, very thick,text=black] at (9,9) (0) {$S_1$};
        \end{tikzpicture}
        \caption{Graph view from $c\mathcal{N}_2$}
        \label{fig:Raptorc_N2}
    \end{subfigure}
 \hspace{7mm}
    \begin{subfigure}[h]{.3\textwidth}
        \centering
        \begin{tikzpicture}[scale=.47, transform shape]
            \color{black}
            \fill[fill=gray, fill opacity=0.2] (1,0) rectangle (3,10) node[text=black, opacity=1,above] {$\mathcal{U}$};
            \filldraw[color=red!60, fill=red!5, very thick](2,1) circle (0.75) node[text=black, opacity=1] {$U_5$};
            \filldraw[color=red!60, fill=red!5, very thick](2,3) circle (0.75) node[text=black, opacity=1] {$U_4$};
            \filldraw[color=red!60, fill=red!5, very thick](2,5) circle (0.75) node[text=black, opacity=1] {$U_3$};
            \filldraw[color=red!60, fill=red!60, very thick](2,7) circle (0.75) node[text=black, opacity=1] {$U_2$};
            \filldraw[color=red!60, fill=red!60, very thick](2,9) circle (0.75) node[text=black, opacity=1] {$U_1$};    
            \draw [-] (2.75,5) to (8.25,9);
            \draw [-] (2.75,5) to (8.25,3);
            \draw [-] (2.75,5) to (8.25,1);
            \draw [-] (2.75,3) to (8.25,7);
            \draw [-] (2.75,3) to (8.25,3);
            \draw [dotted,-] (2.75,3) to (8.25,1);
            \draw [dotted,-] (2.75,1) to (8.25,5);
            \draw [dotted,-] (2.75,1) to (8.25,1);
            \fill[fill=gray, fill opacity=0.2] (8,0) rectangle (10,10) node[text=black, opacity=1,above] {$\mathcal{S}$};
            \node[square,color=blue!60, fill=blue!5, very thick,text=black] at (9,1) (0) {$S_5$};
            \node[square,color=blue!60, fill=blue!5, very thick,text=black] at (9,3) (0) {$S_4$};
            \node[square,color=blue!60, fill=blue!5, very thick,text=black] at (9,5) (0) {$S_3$};
            \node[square,color=blue!60, fill=blue!5, very thick,text=black] at (9,7) (0) {$S_2$};
            \node[square,color=blue!60, fill=blue!5, very thick,text=black] at (9,9) (0) {$S_1$};
        \end{tikzpicture}
        \caption{Graph view from BS}
        \label{fig:Raptorc_BS}
\end{subfigure}
\caption{Bipartite graphs-based {message decoding} received from the main links $\mathcal{G}_M$, connecting node $c\mathcal{N}_1$, and $c\mathcal{N}_2$.}
\label{fig:Raptorc_Graph}
\end{figure}

{The pseudo-code for the adopted} graph-based decoding algorithm {is described in the Algorithm \ref{algo:bipartite}. Moreover, Fig. \ref{fig:Raptorc_Graph} depicts an} {additional}  graph from BS $\mathcal{G}_{\text{BS}}$ and two $c\mathcal{N}$s ($\mathcal{G}_{c \mathcal{N}_1}$ and $\mathcal{G}_{c \mathcal{N}_2}$) seen by DCU.

\begin{algorithm}
\SetAlgoLined
\KwData{Connected bipartite graphs $\mathcal{G}_{\text{BS}}$ and $\mathcal{G}_{c \mathcal{N}_j}$.}
\KwResult{Information $m$  decoded.}
\For{$j=1$ until $q$}{
\textbf{Loop Passive Nodes $\mathcal{G}_{c \mathcal{N}_j}$:} Analyze $\mathcal{G}_{c \mathcal{N}_j}$ and access user $U_i$ connected to a slot node having degree $d=1$;  
\If{degree $d = 1$}{
    Obtain information of $U_i$ \;
    Deduct the collected signals in all slot nodes correlated  to user $U_i$ with user $U_i$'s information at graph $\mathcal{G}_{c \mathcal{N}_j}$ \;
    Send information of $U_i$ via $S_{c \mathcal{N}_j}$ and deduct all slot nodes correlated to user $U_i$ at graph $\mathcal{G}_{\text{BS}}$\;
    }
    }
    \textbf{Loop BS $\mathcal{G}_{\text{BS}}$:} Implement \gls{SIC} decoding algorithm as describe in Section \ref{sec:SIC};\
 \caption{{\bf Network Decoding with RapC }}\label{algo:bipartite}
\end{algorithm}

\vspace{2mm}
\noindent{\bf SP-\gls{IRSA} \& SP-RapIRSA}. \gls{SP-IRSA}  and Service Priority-based Raptor Code Irregular Repetition Slotted ALOHA (SP-RapIRSA) protocols employ additional preamble information to denote the number of packets in the channel, thus the high priority services have higher repetition probability. {Such approach based on service priority} does not consider the access probabilities adjusted according to the traffic load as Irregular Repetition Slotted ALOHA with Priority (P-IRSA) \cite{Sun2016}, which requires a further access controller. SP-\gls{IRSA} protocol arranges $m$ users  into $P_L$ different priority levels based on service-defined priority. Each user carrier services ID to determine whether repetition rate changes according to priority. The categorical priority level vector $\vec{p}_L$ increases (or decreases) the repetition probability of  $\Lambda_\text{d}$. With a finite number of SG applications, {the level} $P_L$ of each service modifies the maximum of $d_\text{m}$ replicas a user $m$ {could send}. Consequently, a user with a higher-level priority sends more replicas of its packet.

\vspace{2mm}
\noindent{\bf SP-S-ALOHA}. Following the SP-\gls{IRSA} description, we have implemented the same Service Priority-based on the S-ALOHA algorithm, namely SP-S-ALOHA. In this case, the service priority modifies the backoff maximum value in the user connection. Whether if the backlogged active users no comply in the attempt, the \gls{QoS} system counts as an error, {\it i.e.}, in the worst-case scenario where backoff value is close to the maximum service priority, {the} user only will {have} an attempt to complete the uplink connection.

\section{Numerical Results} \label{sec:results}
In this section, we present the simulation results of the proposed service priority RA algorithms based on S-ALOHA, IRSA, and \gls{RapIRSA} protocols. The simulation process directly follows the \gls{SG} RA protocols revisited in Section \ref{sec:RA-SGC}, as well as the proposed RA algorithms for \gls{m-SGC} networks described in Section \ref{sec:Proposed_RA}. All the simulations use the \textsc{Matlab} and \textsc{Julia} programming languages. The main parameter values adopted in numerical simulations are summarized in Table \ref{tab:sim_para}. We consider a future AMI  scenario {of \gls{SG} networks,} an urban area with up to {$10^{6}$} SM uniformly distributed in a {$1\times1$ km} service area. The coordinator node is center-located, and the $c\mathcal{N}$s, corresponding to the \gls{RapIRSA} protocols are located to access as many as possible different devices.

\begin{table}[htbp!]
\centering
\caption{Simulation Parameters}
\label{tab:sim_para}
\begin{tabular}{p{3.6cm}p{4.5cm}}
\hline
\textbf{Parameter} & \textbf{Value} \\ 
\hline
{Service Area} & {$A=1 \text{km}^2$}\\
\ta{\# Active users, $m$} 
& Up to $M=10^{6}$ devices\\
\ta{Network Load} & \ta{$G=\frac{m}{n}  \in \,\, ]0;\,1.2]  \,\,\,\left[\frac{\text{user}}{\text{slot}}\right]$}\\
Active Users {distribution} & Poisson \& Beta \\
{Time-slot duration} & {$1$ ms} \\
\gls{IRSA} \gls{RAF} Length & $ n_{{\textsc{raf}}}=50$ \textit{slots}\\ 
\# Replicas (Max.) &  $d_\text{m} = 8$ \\ 
Degree distribution  & $\Lambda_8 (x)=0.5x^2 + 0.28x^3 + 0.22x^8 $\\ 
\# \gls{SIC} iterations (Max.) & ${\mathcal{I}=}\, 20$ \\
 \hline
{S-ALOHA} Back-off limit & $B_{\rm off}=50$ \textit{slots} \\ 
\hline
number of $c\mathcal{N}$ & $q = [2 \quad 3 \quad 8]$ \\
{fraction} of 
$n_{{\rm RapC}}$ & $\eta = 0.25$ \\
 \hline
{\# SP $m$ Users} & Up to $M= 10^{4}$ devices \\
Priority Level & $\vec{p}_L=[0 \text{(max)} \quad 100 \text{(min)}]$  \\ \hline
\#  \glssymbol{MCS} Realizations & $100$ \\
Simulation time & $10$ s \\ 
 \hline
\end{tabular}%
\end{table}
 
\begin{figure}[htbp]
\centering
\includegraphics[trim=9mm 1mm 10mm 8mm,clip, width=.7\textwidth]{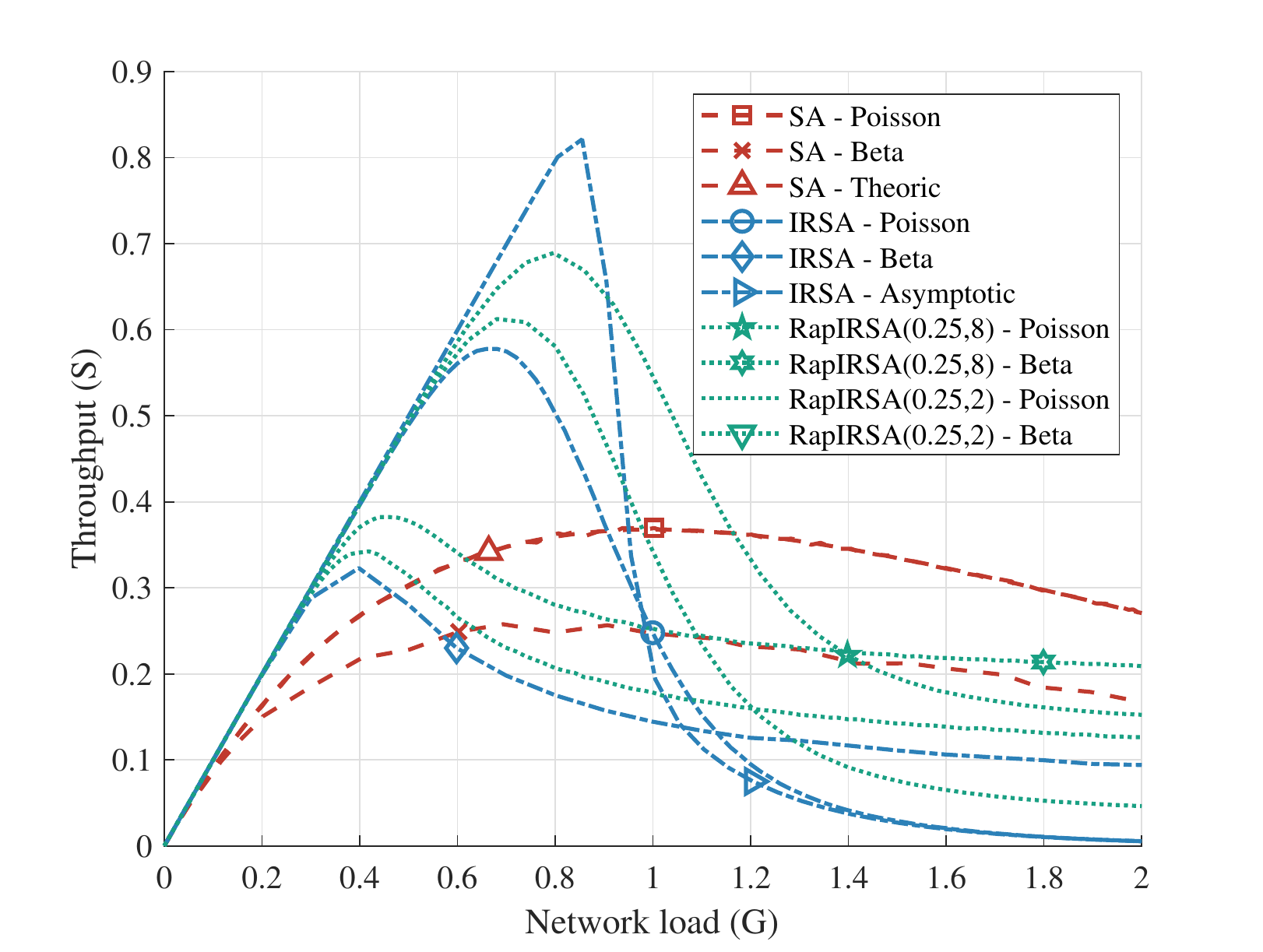}
\caption{Simulated {throughput} for SA, and for \gls{IRSA} and \gls{RapIRSA} with $\Lambda_8(x)$.} 
\label{fig:t_v_l}
\end{figure}

Fig. \ref{fig:t_v_l} {depicts} the throughput for S-ALOHA (SA),  IRSA, RapIRSA$(0.25,8)$, and RapIRSA$(0.25,2)$ implementations for two different users' arrival distributions. This simulation includes the theoretical throughput for S-ALOHA algorithm considering the Poisson distribution and an asymptotic ($n_{\textsc{raf}} \rightarrow \infty$) representation of the \gls{IRSA} protocol. Note that \gls{RapIRSA} improved throughput not only for $G>1$ but its performance is also comparable with the asymptotic \gls{IRSA} with the Poisson distribution. It's worth mentioning, \gls{RapIRSA} for  $G>1.4$ present a better throughput under traffic governed by a Beta distribution. In our specific scenario, this finding represents a breakthrough to determine that the \gls{RapIRSA} protocol attains better performance \ta{operating under overloaded networks}.

\begin{figure}[htbp]
    \centering
    \includegraphics[width=.7\textwidth]{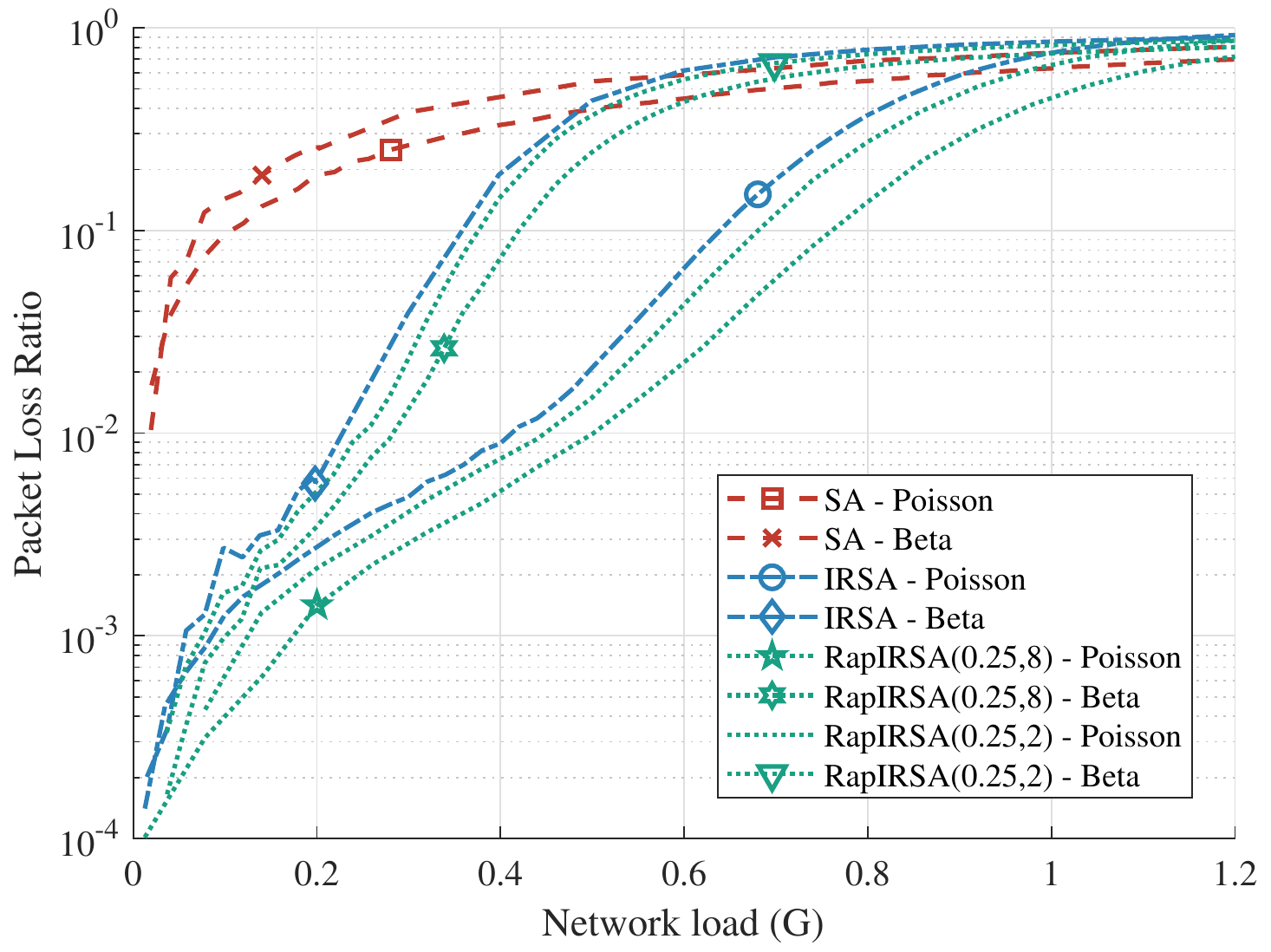}
    \caption{Packet loss ratio for SA, and for \gls{IRSA} with $\Lambda_8 =0.5x^2 + 0.28x^3 + 0.22x^8$.} 
    \label{fig:PLR}
    \vspace{-2mm}
\end{figure}

Accordingly, Fig. \ref{fig:PLR} depicts the packet loss ratio versus the network load. Under {low and moderate-to-high} network loading condition, {{\it i.e.}, $0<G<1.2$ (Poisson distribution) and $0<G<0.6$ (Beta distribution),} those values indicate a considerable advantage from \gls{IRSA} and \gls{RapIRSA} over S-ALOHA protocol, considering both arrival distributions. \(\Lambda_8(x)\) is a commonly used degree distribution in the literature for modeling various types of networks, including communication networks and power grids. It is a power law distribution, which means that it has a heavy tail that captures the presence of a small number of highly connected nodes in the network. This makes it a suitable choice for modeling networks with heterogeneous connectivity patterns, such as smart grid networks. Moreover, the choice of \(\Lambda_8(x)\) is consistent with previous work on modeling smart grid networks, which has also used power law degree distributions \cite{Bolognani2013, Tse2018}.

\subsection{Latency}
\begin{figure}[htbp]
\centering
\includegraphics[width=.75\textwidth]{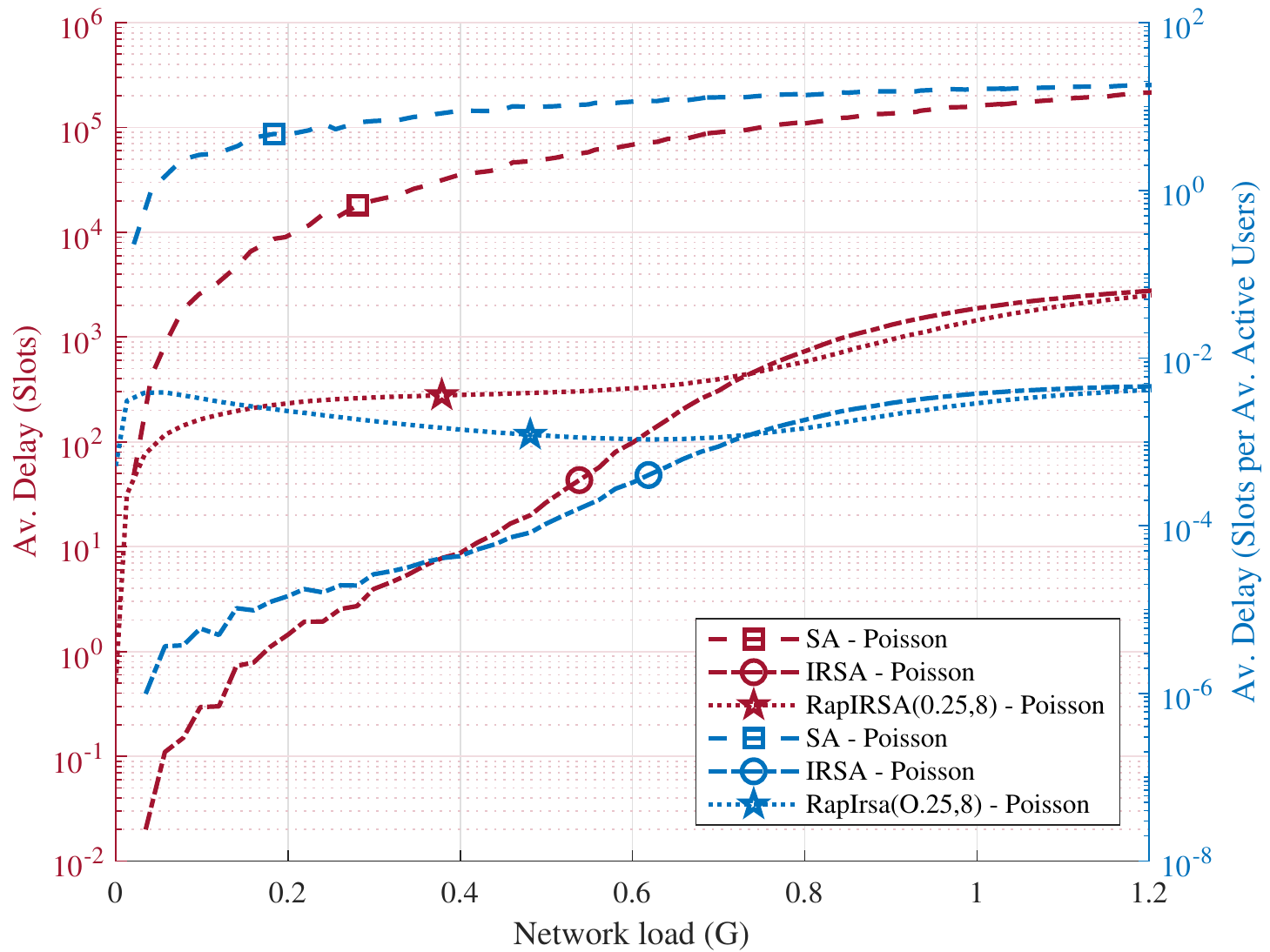}
\caption{Average delay in terms of  \# slots (left y-axis) and  \# slots per average \# active users (right y-axis) {\it vs} Network loading.}
\label{fig:Ave_Delay}
\end{figure}

When we first introduced Fig. \ref{fig:RapIRSA_time} the extra slots $n_q$ for each frame in the \gls{RapIRSA} protocol, it creates concern about increasing latency. As delay constraints are fundamental to achieve \gls{QoS} requirements, an additional evaluation of these parameters is presented in Fig. \ref{fig:Ave_Delay} to chart those values within \gls{IRSA} and SA protocols. Fig. \ref{fig:Ave_Delay}  shows that the \ta{{\it average delay} [slots] (red curves) and the associated {\it average delay} [slots/active users] (blue curves),}  for all investigated algorithms. Hence, as network loading increases, S-ALOHA maintains a steadily increasing behavior, resulting in high delays (over $10^5$ slots) over high network loading ranging from $1<G<2$. Moreover, the \gls{IRSA} protocol presents a remarkable performance without delay for low-medium network loading ($G<0.5$). Furthermore, \gls{RapIRSA} keeps an almost constant operation point for all network loading. Even for measurement of \ta{the average delay (slots) per active users} in Fig. \ref{fig:Ave_Delay} (right y-axis), the delays of the S-ALOHA protocol  appear unfeasible under requirements described in Table \ref{tab:latency} for many of the \gls{SG} applications. {Table \ref{tab:latency} scales priority in \gls{SG} application type based on maximum tolerable latency.} 
 
\vspace{-2mm}

\subsection{Application Complying \ta{Ratio} (ACR) {Reliability}}

\begin{figure}[htbp]
\centering
\includegraphics[trim=4mm .3mm 7mm 6mm,clip,width=.7\textwidth]{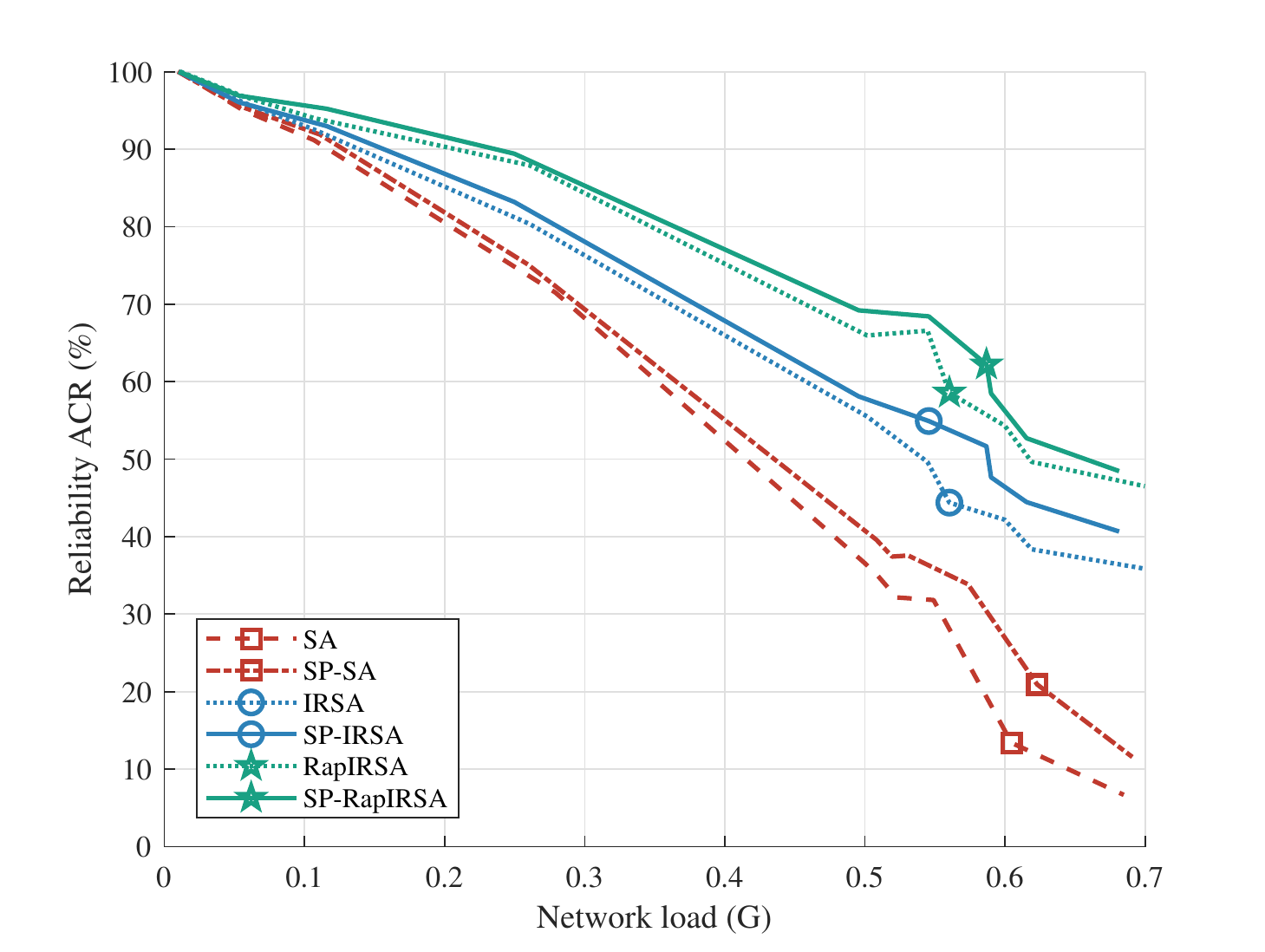}
\caption{Application complying rate (\%) to measure algorithm capability to achieve \gls{QoS} requirement.}
\label{fig:ACR}
\end{figure}

\color{black}
Fig. \ref{fig:ACR} compares the percentage {of ACR reliability}  to {evaluate the RA} algorithm capability {in achieving a certain} \gls{QoS} requirement. {Under such specific reliability metric, and the modifications in the IRSA protocol introduced in Sec. \ref{sec:Proposed_RA}, which are based on Raptor codes deployment and service priority, the \gls{QoS} information was labeled into the active user's preamble}. {The ACR reliability metric is evaluated with up to $10^4$ devices; all SP RA protocols discussed in this work use the $\vec{p}_L$ values described in Table \ref{tab:latency}.} The {ACR reliability curves in Fig. \ref{fig:ACR}}  corroborate the viability of the proposed {RapIRSA and SP-RapIRSA} algorithms to improve RA protocols performance, as well as, reach \gls{QoS} requirements, described as one of the main objectives of the proposed work. Both \gls{RapIRSA} variants, SP-\gls{RapIRSA} and RapIRSA, provide the best ACR values across the entire networking loading {range}. {Moreover,} Fig. \ref{fig:ACR} {reveals that the critical scenario of $90\%$ ACR reliability ($\mathcal{R}_{\text{ACR}}=0.9$), is attained for different network loading, depending of which RA protocol is adopted; the best performance is achieved by the SP-RapIRSA for a network loading of $G=0.2$, followed by the RapIRSA protocol}. The \gls{ACR} reliability metric reveals the crucial trade-off between a delivered packet within the latency constraints. {Notice that promising RA protocols for m-SGC applications require considering carefully the ACR reliability constraint.}

\vspace{-1mm}

\ta{Combining the superior ACR performance depicted in Fig. \ref{fig:ACR} with the reduced packet loss ratio (Fig. \ref{fig:PLR}), balanced delay {\it vs.} network load in Fig. \ref{fig:Ave_Delay}, and suitable throughput (Fig. 8) attained by the \gls{RapIRSA} protocol,} one can infer that among the three \gls{GF} RA protocols analyzed, the \gls{RapIRSA} \ta{is promising and} the most feasible protocol for \gls{SG} applications. Even with the addition of extra slots for the $c\mathcal{N}$ delays, \gls{RapIRSA} can comply with tied throughput, PLR, and latency performance requirements in the current \gls{SG} networks.

\subsection{{Computational Load and Memory Requirements}} \label{sec:comp_load}
{The computational load and memory requirements imposed by the proposed RA algorithms are evaluated by simulations performed on a workstation with an Intel Core i7-10700K CPU, 3.70GHz, and 16GB of RAM.
Table \ref{tab:comp_load} summarizes the computational load in terms of the average processing time per packet and memory requirements of the RA algorithms. The {\it processing time} was measured for the analyzed RA algorithms based on S-ALOHA, IRSA, and \gls{RapIRSA} protocols, respectively. The computational load of the proposed RA algorithms based on IRSA and \gls{RapIRSA} protocols is substantially higher than the S-ALOHA protocol. Specifically, the  \gls{RapIRSA} presents a higher computational load than IRSA due to the additional processing time required for the encoding and decoding process of Raptor codes. However, the difference in processing time is negligible for a single packet. The memory requirement of the proposed RA algorithms is also presented in Table \ref{tab:comp_load}. As expected, the memory requirement of the \gls{RapIRSA} protocol is higher than IRSA ($2\times$) and S-ALOHA ($20\times$)  due to the additional storage requirement for encoded packets.}

\begin{table}[!htbp]
\caption{{Computational Load and Memory Requirements of RA Algorithms.}}
\label{tab:comp_load}
\centering
{
\begin{tabular}{ccc}
\hline
\bf RA Algorithm &\bf  Avg. Processing Time/Packet (ms) & \bf Memory Requirement (KB) \\
\hline
S-ALOHA & 0.1 & 1 \\
IRSA & 0.5 & 10 \\
\gls{RapIRSA} & 0.6 & 20 \\
\hline
\end{tabular}
}
\end{table}

{Overall, the proposed IRSA and \gls{RapIRSA} RA algorithms require higher computational load and memory compared to the S-ALOHA protocol. However, considering the higher reliability and lower packet loss rate achieved by the proposed RA algorithms, the additional computational load and memory requirement can be justified, both attaining improvements in the performance-complexity trade-off. Finally, the computational load and memory requirement of both RA protocols can be further optimized by employing efficient encoding and decoding techniques for Raptor codes.}

\section{Conclusions} \label{sec:conclusion}
We propose a service priority RA protocol for \gls{SG} communication network applications. The proposed algorithm is based on the S-ALOHA and \gls{IRSA} RA protocols with adapting parameters considering the \gls{QoS} requirements as priority, latency, and data rates metrics. The proposed protocol aims to enhance the probability of success of various \gls{SG} applications. 

The \gls{RapIRSA} protocol is proposed in this paper to guarantee better throughput of active users with the help of connecting nodes ($c\mathcal{N}$). We have shown that relative to SA and \gls{IRSA} in particular for multipacket messages, the \gls{RapIRSA} can increase the reliability and keep the throughput above all others. The \gls{RapIRSA} is better than SA and IRSA, indicating this protocol is suitable for \gls{SG} applications.

\ta{The investigation of adapting \gls{IRSA} \& \gls{RapIRSA} for smart sensors' future applications can be guided by the following issues and findings: 
{\bf a}) such protocols} reach their peak performance without latency constraints; 
\ta{{\bf b})} they rely on reliability constraints rather than latency constraints;
\ta{{\bf c})} latency constraints on  \gls{IRSA} \& \gls{RapIRSA} depend on the limit of the frame size; nevertheless, a decreasing frame size supports fewer users, rejecting some users from the system; \ta{{\bf d})} \gls{IRSA} \& \gls{RapIRSA} represent acceptable solutions for many critical \ta{applications}.



\end{document}

%% file: Acronyms.tex
\newacronym{mSG}{$\mu$SG}{micro-SG}
\newacronym{1PPS}{1PPS}{One pulse-per-second}
\newacronym{3GPP}{3GPP}{3rd Generation Partnership Project}
\newacronym{ADC}{ADC}{Analog-to-Digital conversion}
\newacronym{AI}{AI}{Artificial Intelligence}
\newacronym{AMI}{AMI}{Advanced Metering Infrastructure}
\newacronym{ANN}{ANN}{Artificial Neural Networks}
\newacronym{BLER}{BLER}{Block Error rate}
\newacronym{CB}{CB}{Circuit Breaker}
\newacronym{CBG}{CBG}{Code Block Group}
\newacronym{CES}{CES}{Conventional Energy Systems}
\newacronym{CIS}{CIS}{Consumer Information System}
\newacronym{CP}{CP}{Cycle Prefix}
\newacronym{CPU}{CPU}{Central Process Unit}
\newacronym{CT}{CT}{Current Transformer}
\newacronym{DA}{DA}{Distribution Automation}
\newacronym{DER}{DER}{Distributed Energy Resources}
\newacronym{DFR}{DFR}{Digital Fault Recording}
\newacronym{DFT}{DFT}{Discrete Fourier Transform}
\newacronym{DG}{DG}{Distributed Generation}
\newacronym{DM}{DM}{Distribution Management System}
\newacronym{DMS}{DMS}{Distribution Management System}
\newacronym{DoS}{DoS}{Denial of Service}
\newacronym{DR}{DR}{Demand Response}
\newacronym{DSL}{DSL}{Digital Subscriber Line}
\newacronym{DSM}{DSM}{Demand Side Management}
\newacronym{DSO}{DSO}{Distribution System Operators}
\newacronym{DSP}{DSP}{Digital Signal Processors}
\newacronym{DWT}{DWT}{Discrete Wavelet Transform}
\newacronym{E2E}{E2E}{End-to-End}
\newacronym{EE}{EE}{Energy-Efficient}
\newacronym{EMS}{EMS}{Energy Management System}
\newacronym{eMTC}{eMTC}{Enhanced Machine-Type Communication}
\newacronym{EV}{EV}{Electrical Vehicle}
\newacronym{FAN}{FAN}{Field Area Network}
\newacronym{FDLSG}{FD/L-SG}{Fault Detection and/or Location in SG Systems}
\newacronym{FFT}{FFT}{Fast Fourier transform}
\newacronym{GPS}{GPS}{Global Positioning System}
\newacronym{H2H}{H2H}{Human-to-Human}
\newacronym{HAN}{HAN}{Home Area Network}
\newacronym{HMI}{HMI}{Human Machine Interface}
\newacronym{IAB}{IAB}{Integrated Access Backhaul}
\newacronym{ICT}{ICT}{Information and Communication Technology}
\newacronym{IED}{IED}{Intelligent Electronic Device}
\newacronym{IGBT}{IGBT}{Insulated-Gate Bipolar Transistor}
\newacronym{IoE}{IoE}{Internet-of-Energy}
\newacronym{IoT}{IoT}{Internet-of-things}
\newacronym{IoTG}{IoTG}{Internet-of-things-Grid}
\newacronym{IP}{IP}{Internet Protocol}
\newacronym{IT}{IT}{Information Technology}
\newacronym{IRIGB}{IRIG-B}{Inter-Range Instrumentation Group-B}
\newacronym{ISP}{ISP}{Internet Service Providers}
\newacronym{LoS}{LoS}{Line of Sight}
\newacronym{LRWPAN}{LR-WPAN}{Low-Rate Wireless Personal Area Networks}
\newacronym{LTE}{LTE}{Long-Term Evolution}
\newacronym{M2M}{M2M}{Machine-to-machine}
\newacronym{MA}{MA}{Moving Average}
\newacronym{MDMS}{MDMS}{Meter Data Management System}
\newacronym{ML}{ML}{Machine Learning}
\newacronym{mMTC}{mMTC}{massive Machine-Type Communication}
\newacronym{MPU}{MPU}{Main Processing Unit}
\newacronym{MTC}{MTC}{Machine-Type Communication}
\newacronym{MU}{MU}{Merging Units}
\newacronym{NAN}{NAN}{Neighborhood Area Network}
\newacronym{NBIoT}{NB-IoT}{Narrowband Internet-of-Things }
\newacronym{NN}{NN}{Neural Network}
\newacronym{NSP}{NSP}{Network Service Providers}
\newacronym{NTP}{NTP}{Network Time Protocol}
\newacronym{OPC}{OPC}{Open Platform Communications}
\newacronym{P2G}{P2G}{Phase to Ground}
\newacronym{P2P}{P2P}{Single-Phase to Phase}
\newacronym{PDC}{PDC}{Phasor Data Concentrator}
\newacronym{PDCP}{PDCP}{Packet Data Convergence Protocol}
\newacronym{PEM}{PEM}{Proton Exchange Membrane}
\newacronym{PLCo}{PLC}{Programmable Logic Controller}
\newacronym{PLC}{PLC}{Power Line Communication}
\newacronym{PMU}{PMU}{Phasor Measurement Units}
\newacronym{PP2G}{PP2G}{Two-Phase to Ground}
\newacronym{PPP}{PPP}{Three-Phase}
\newacronym{PPP2G}{PPP2G}{Three-Phase to Ground}
\newacronym{PQA}{PQA}{Power Quality Analyzers}
\newacronym{PSFB}{PSFB}{Phase-Shifted Full-Bridge}
\newacronym{PTP}{PTP}{Precision Time Protocol}
\newacronym{PV}{PV}{Photovoltaic}
\newacronym{PWM}{PWM}{Pulse Width Modulation}
\newacronym{QoS}{QoS}{Quality of Service}
\newacronym{RA}{RA}{Random Access}
\newacronym{RACH}{RACH}{Random Access Channel}
\newacronym{RAN}{RAN}{Radio Access Network}
\newacronym{RapIRSA}{RapIRSA}{Combining Raptor Codes and IRSA}
\newacronym{RF}{RF}{Radio frequency}
\newacronym{RIAPS}{RIAPS}{Resilient Information Architecture Platform for Smart Grid}
\newacronym{RTU}{RTU}{Remote Terminal Units}
\newacronym{SA}{SA}{Service and System Aspects}
\newacronym{SCADA}{SCADA}{Supervisory Control and Data Acquisition}
\newacronym{SDHSONET}{SDH/SONET}{Synchronous Digital Hierarchy/Synchronous Optical Network}
\newacronym{SG}{SG}{Smart Grids}
\newacronym{SGSC}{SGSC}{Smart Grid Security Classification}
\newacronym{SM}{SM}{Smart Meters}
\newacronym{SMPS}{SMPS}{Synchronized Phasor Measurement System}
\newacronym{SPM}{SPM}{Synchro-Phasor Measurement}
\newacronym{SS}{SS}{Smart Sensor}
\newacronym{ST}{ST}{Smart Transformer}
\newacronym{SV}{SV}{Sampled Values}
\newacronym{SVM}{SVM}{Support Vector Machine}
\newacronym{SyG}{SyG}{Synchronous Generators}
\newacronym{TSC}{TSC}{Time-Sensitive Communications}
\newacronym{TSN}{TSN}{Time-Sensitive Networks}
\newacronym{URLLC}{URLLC}{Ultra-Reliable Low-Latency Communications}
\newacronym{VSC}{VSC}{Voltage-Sourced Converter}
\newacronym{VT}{VT}{Voltage Transformer}
\newacronym{WAMS}{WAMS}{Wide Area Measurement System}
\newacronym{WAN}{WAN}{Wide Area Network}
\newacronym{WASA}{WASA}{Wide-Area Situational Awareness}
\newacronym{WiMAX}{WiMAX}{Worldwide Interoperability for Microwave Access}
\newacronym{WLAN}{WLAN}{Wireless Local Area Networks}
\newacronym{LPWAN}{LPWAN}{low-power wide-area network }
\newacronym{VoIP}{VoIP}{Voice over IP}
\newacronym{WSN}{WSN}{wireless sensor network}

\newacronym{GB}{GB}{{\it grant-based}}
\newacronym{GF}{GF}{{\it grant-free}}
\newacronym{SIC}{SIC}{successive interference cancellation }
\newacronym{CRDSA}{CRDSA}{contention resolution diversity slotted  ALOHA }
\newacronym{IRSA}{IRSA}{irregular repetition slotted ALOHA}
\newacronym{Saloha}{S-ALOHA}{slotted-ALOHA}
\newacronym{m-SGC}{m-SGC}{massive smart grid communication}
\newacronym{SGC}{SGC}{smart grid communication}
\newacronym{BS}{BS}{base station}
\newacronym{RAF}{RAF}{Random Access Frame}
\newacronym{MCS}{MCS}{Monte Carlo Simulations}

\newacronym{SUN}{SUN}{smart utility network}
\newacronym{CSMA}{CSMA}{Carrier-sense Multiple Access}
\newacronym{VVWC}{VVWC}{Voltage, Var, and Watt Control}
\newacronym{ADR}{ADR}{automated demand response}
\newacronym{DLR}{DLR}{Dynamic Line Rating}
\newacronym{DS}{DS}{distributed storage }
\newacronym{DCU}{DCU}{data collector unit }
\newacronym{PDR}{PDR}{Packet Delivery Rate}
\newacronym{PLR}{PLR}{Packet Loss \ta{Ratio}}
\newacronym{FEC}{FEC}{forward error correction}
\newacronym{HD}{HD}{Hard Delay}
\newacronym{SD}{SD}{Soft Delay}
\newacronym{IC}{IC}{interference canceller}
\newacronym{LDPC}{LDPC}{low-density parity-check}
\newacronym{BEC}{BEC}{binary erasure channel}
\newacronym{SDOs}{SDOs}{Standards Development Organizations}
\newacronym{ScS}{ScS}{Subcarrier-Spacing}
\newacronym{NR}{NR}{New Radio}
\newacronym{SP-IRSA}{SP-IRSA}{Service Priority-based Irregular Repetition Slotted ALOHA}
\newacronym{MPR}{MPR}{multipath-relaying}
\newacronym{ARQ}{ARQ}{automatic repeat request}
\newacronym{MPckR}{MPckR}{multipacket reception}
\newacronym{RapC}{RapC}{Raptor codes-structured}
\newacronym{SN}{SN}{slot number}
\newacronym{ACR}{ACR}{Application Complying \ta{Ratio}}
\newacronym{UL}{UL}{Up-Link}
\newglossaryentry{a}{type=symbols,name=$A$,description={service area}}
\newglossaryentry{boff}{type=symbols,name=$B_{{\rm off}}$,description={S-Aloha Back-off limit}}
\newglossaryentry{bwreq}{type=symbols,name=$BW_{{\rm req}}$,description={required network bandwith}}
\newglossaryentry{dm}{type=symbols,name=$d_m$,description={maximum number of replicas} }
\newglossaryentry{d}{type=symbols,name=$d$,description={repetition rate}}
\newglossaryentry{feq}{type=symbols,name=$f$,description={utility frequency}}
\newglossaryentry{f}{type=symbols,name=$F$,description={frame}}
\newglossaryentry{G}{type=symbols,name=$G$,description={channel traffic or load}}
\newglossaryentry{lambdaf}{type=symbols,name=$\Lambda(x)$,description={user degree distribution}}
\newglossaryentry{lamda}{type=symbols,name=$\lambda$,description={parameter of the Poisson distribution}}
\newglossaryentry{lc}{type=symbols,name=$\Delta_\text{C}$,description={critical Latency Response}}
\newglossaryentry{ln}{type=symbols,name=$\Delta_\text{N}$,description={network latency}}
\newglossaryentry{lx}{type=symbols,name=$\Delta_\text{X}$,description={network changing state latency}}
\newglossaryentry{l}{type=symbols,name=$L$,description={latency}}
\newglossaryentry{mv}{type=symbols,name=$M_v$,description={r.v. of the active users}}
\newglossaryentry{m}{type=symbols,name=$m$,description={active users in at any time $t$/frame}}
\newglossaryentry{M}{type=symbols,name=$M$,description={number of nodes}}
\newglossaryentry{nraf}{type=symbols,name=$n_{{\textsc{raf}}}$,description={slots of an IRSA frame $RAF$}}
\newglossaryentry{n}{type=symbols,name=$n$,description={slot}}
\newglossaryentry{pdr}{type=symbols,name=$PDR$,description={packet delivery ratio}}
\newglossaryentry{pgi}{type=symbols,name=$P_{\text{G}_i}$,description={number of packets generated by user $i$}}
\newglossaryentry{pg}{type=symbols,name=$P_\text{G}$,description={number of packets generated}}
\newglossaryentry{pks}{type=symbols,name=$pk_{\rm s}$,description={application packet size}}
\newglossaryentry{plr}{type=symbols,name=$PLR$,description={packet loss ratio}}
\newglossaryentry{pri}{type=symbols,name=$P_{R_i}$,description={number of packets received by user $i$}}
\newglossaryentry{pr}{type=symbols,name=$P_\text{R}$,description={number of packets received}}
\newglossaryentry{r}{type=symbols,name=$\mathcal{R}$,description={reliability}}
\newglossaryentry{racr}{type=symbols,name=$\mathcal{R}_\text{ACR}$,description={ACR reliability}}
\newglossaryentry{sf}{type=symbols,name=$SF$,description={sub-frame}}
\newglossaryentry{s}{type=symbols,name=$S$,description={throughput}}
\newglossaryentry{tac}{type=symbols,name=$T_{\rm Ac}$,description={accumulation time}}
\newglossaryentry{tauhd}{type=symbols,name=$\tau_{\rm HD_{req}}$,description={HD delay requirement}}
\newglossaryentry{taureq}{type=symbols,name=$\tau_{\rm req}$,description={delay requirement}}
\newglossaryentry{tau}{type=symbols,name=$\tau$,description={delay factor}}
\newglossaryentry{ta}{type=symbols,name=$T_{\rm A}$,description={activation time}}
\newglossaryentry{tf}{type=symbols,name=$T_\text{F}$,description={duration of a frame $F$}}
\newglossaryentry{traf}{type=symbols,name=$T_{\rm RAF}$,description={duration of an IRSA frame $RAF$}}
\newglossaryentry{tsf}{type=symbols,name=$T_\text{SF}$,description={duration of a sub-frame $SF$}}
\newglossaryentry{ts}{type=symbols,name=$T_\text{S}$,description={duration of a slot}}
\newglossaryentry{tw}{type=symbols,name=$T_\text{W}$,description={waiting time}}
\newglossaryentry{t}{type=symbols,name=$T$,description={electrical utility cycle}}
\newglossaryentry{cn}{type=symbols,name=$c\mathcal{N}$,description={connecting nodes}}

\newglossaryentry{snfr}{type=symbols,name=$n_{\rm RapC}$,description={SN in the frame}}

\newglossaryentry{ncn}{type=symbols,name={$q$},description={number of $c\mathcal{N}$}}

\newglossaryentry{eta}{type=symbols,name=$\eta$,description={ target throughput coefficient}}

\newglossaryentry{bgs}{type=symbols,name=$n_q$,description={{SN} of {SN of $c\mathcal{N}$s}}}

\newglossaryentry{graphm}{type=symbols,name=$\mathcal{G}_{\text{BS}}$,description={SN of Graph as seen in BS}}

\newglossaryentry{ustotal}{type=symbols,name=$U_m$,description={number of total active users during $N$ slots}}

\newglossaryentry{utreq}{type=symbols,name=$U_i^k$,description={active users $i$ who succeed in sending a packet in each $k$ slot within the latency requirement $\tau_{{\rm req}_i}$}}